\documentclass[aps,prl,twocolumn,groupedaddress]{revtex4-1}
\usepackage{graphicx}
\usepackage{caption}
\DeclareCaptionLabelSeparator{dot}{. }
\makeatletter
\def\justified{
	\let\\\@normalcr
	\@rightskip\z@skip \rightskip\@rightskip
	\leftskip\z@skip
	\parindent 0em\relax
	\setlength{\parfillskip}{0pt plus 1fil}}
\DeclareCaptionJustification{justified}{\justified}
\usepackage{subcaption}
\usepackage{amsmath}
\usepackage{mathrsfs} 
\usepackage{braket}
\usepackage{units}
\usepackage{ragged2e}
\usepackage[colorlinks,urlcolor=blue ,citecolor=blue ,linkcolor=blue ]{hyperref}
\usepackage{xcolor}
\usepackage{nicefrac} 
\usepackage{lipsum}
\usepackage{nameref}
\usepackage{hyperref}
\usepackage{textcomp} 
\usepackage{urwchancal}
\usepackage{xcolor}
\usepackage{float}
\definecolor{darkgreen}{rgb}{0,0.5,0}

\newcommand{\bs}{\boldsymbol}

\newcommand{\Er}{\ensuremath{^{166}}{\rm Er} }
\newcommand{\Dy}{\ensuremath{^{164}}{\rm Dy} }
\newcommand{\Dyo}{\ensuremath{^{162}}{\rm Dy} }

\newcommand{\as}{\ensuremath{a_{\rm s}}}
\newcommand{\add}{\ensuremath{a_{\rm dd}}}
\newcommand{\edd}{\ensuremath{\epsilon_{\rm dd}}}
\newcommand{\tho}{t_{\rm h}}
\newcommand{\tr}{t_{\rm r}}

\newcommand{\um}{\mu{\rm m}}

\newcommand{\asf}{\as}

\begin{document}
	
	\title{
Long-lived and transient supersolid behaviors in dipolar quantum gases}
	\author{L.\,Chomaz,$^{1}$ D.\,Petter,$^{1}$ P. Ilzh\"ofer,$^{2}$ G.\,Natale,$^{1}$  A. Trautmann,$^{2}$ C. Politi,$^{2}$ G. Durastante,$^{1,2}$ R.\,M.\,W.\,van\,Bijnen,$^{2}$ A.\,Patscheider,$^{1}$ M.\,Sohmen,$^{1,2}$ M.\,J.\,Mark,$^{1,2}$ and F.\,Ferlaino$^{1,2,*}$}
	
	\affiliation{%
		$^{1}$Institut f\"ur Experimentalphysik, Universit\"at Innsbruck, Technikerstra{\ss}e 25, 6020 Innsbruck, Austria\\
		$^{2}$Institut f\"ur Quantenoptik und Quanteninformation, \"Osterreichische Akademie der Wissenschaften, Technikerstra{\ss}e 21a, 6020 Innsbruck, Austria\\
	}	

	\date{\today}
	
\begin{abstract}
By combining theory and experiments, we demonstrate that dipolar quantum gases of both $\Er$ and $\Dy$ support a state with supersolid properties, where a spontaneous density modulation and a global phase coherence coexist. This paradoxical state occurs in a well defined parameter range, separating the phases of a regular Bose-Einstein condensate and of an insulating droplet array, and is rooted in the roton mode softening, on the one side, and in the stabilization driven by quantum fluctuations, on the other side. 
Here, we identify the parameter regime for each of the three phases. In the experiment, we rely on a detailed analysis of the interference patterns resulting from the free expansion of the gas, quantifying both its density modulation and its global phase coherence. Reaching the phases via a slow interaction tuning, starting from a stable condensate, we observe that $\Er$ and $\Dy$ exhibit a striking difference in the lifetime of the supersolid properties, due to the different atom loss rates in the two systems. Indeed, while in $\Er$ the supersolid behavior only survives a few tens of milliseconds, we observe coherent density modulations for more than 150\,ms in $\Dy$. 
Building on this long lifetime, we demonstrate an alternative path to reach the supersolid regime, relying solely on evaporative cooling starting from a thermal gas. 
\end{abstract}

\maketitle

\section{Introduction}

Supersolidity is a paradoxical quantum phase of matter where both crystalline and superfluid order coexist \cite{Andreev,Chester,Leggett}. Such a counter-intuitive phase, featuring rather antithetic properties, has been originally considered for quantum crystals with mobile bosonic vacancies, the latter being responsible for the superfluid order. Solid $^4$He has been long considered a prime system to observe such a phenomenon \cite{Kirzhnits:1971cco, Schneider:1971}. However, after decades of theoretical and experimental efforts, an unambiguous proof of supersolidity in solid $^4$He is still missing~\cite{Balibar:2010,Boninsegni:2012}.

In search of more favorable and controllable systems, ultracold atoms emerged as a very promising candidate, thanks to their highly tunable interactions. Theoretical works point to the existence of a supersolid ground state in different cold-atom settings, including dipolar \cite{Lu2015sds} and Rydberg particles \cite{Henkel2010tdr, Cinti2010sdc}, cold atoms with a soft-core potential \cite{Boninsegni2012}, or lattice-confined systems \cite{Boninsegni:2012}.  Breakthrough experiments with Bose--Einstein condensates (BECs)  coupled to light have recently demonstrated a state with supersolid properties~\cite{Leonard:2017,Li:2017}. While in these systems indeed two continuous symmetries are broken, the crystal periodicity is set by the laser wavelength, making the supersolid incompressible. 

Another key notion concerns the close relation between a possible transition to a supersolid ground state and the existence of a local energy minimum at large momentum in the excitation spectrum of a non-modulated superfluid, known as roton mode \cite{Landau41}.  Since excitations corresponding to a periodic density modulation at the roton wavelength are energetically favored, the existence of this mode indicates the system's tendency to crystallize ~\cite{Nozieres2004itr} and it is predicted to favor a transition to a supersolid ground state~\cite{Kirzhnits:1971cco, Schneider:1971, Henkel2010tdr}. 

Remarkably, BECs of highly magnetic atoms, in which the particles interact through the long-range and anisotropic dipole-dipole interaction (DDI), appear to gather several key ingredients for realizing a supersolid phase. First, as predicted more than fifteen years ago~\cite{ODell2003,Santos:2003} and recently demonstrated in experiments~\cite{Chomaz:2018, Petter:2018}, the partial attraction in momentum space due to the DDI gives rise to a roton minimum. The corresponding excitation energy, i.\,e.\,the roton gap, can be tuned in the experiments down to vanishing values. Here, the excitation spectrum softens at the roton momentum and the system becomes unstable. Second, there is a non-trivial interplay between the trap geometry and the phase diagram of a dipolar BEC. 
For instance, our recent observations have pointed out the advantage of axially-elongated trap geometries (i.\,e.\,cigar-shaped) compared  to the typically considered cylindrically-symmetric ones (i.\,e.\,
pancake-shaped) 
in enhancing the visibility of the roton excitation in experiments. 
Last but not least, while the concept of a fully softened mode is typically related to instabilities and disruption of a coherent quantum phase, groundbreaking works in the quantum-gas community have demonstrated that quantum fluctuations can play a crucial role in stabilizing a dipolar BEC~\cite{Kadau:2016,Igor:2016,Chomaz:2016,Waechtler:2016,Waechtler:2016b,Schmitt:2016,Igor2018ooa}. Such a stabilization mechanism enables the existence, beyond the mean-field instability, of a variety of stable ground states, from a single macro-droplet \cite{Bisset:2016, Chomaz:2016, Waechtler:2016b} to striped phases \cite{Wenzel:2017}, and droplet crystals~\cite{Baillie:2018}; see also related works~\cite{petrov2015, Cabrera2018, semeghini2018, cheiney2018}. For multi-droplet ground states, efforts have been devoted to understand if a phase coherence among ground-state droplets could be established~\cite{Wenzel:2017, Baillie:2018}. However, previous experiments with $\Dy$ have shown the absence of phase coherence across the droplets~\cite{Wenzel:2017}, probably due to the limited atom numbers.
\begin{figure*}[htbp!]
	\includegraphics[width=\textwidth]{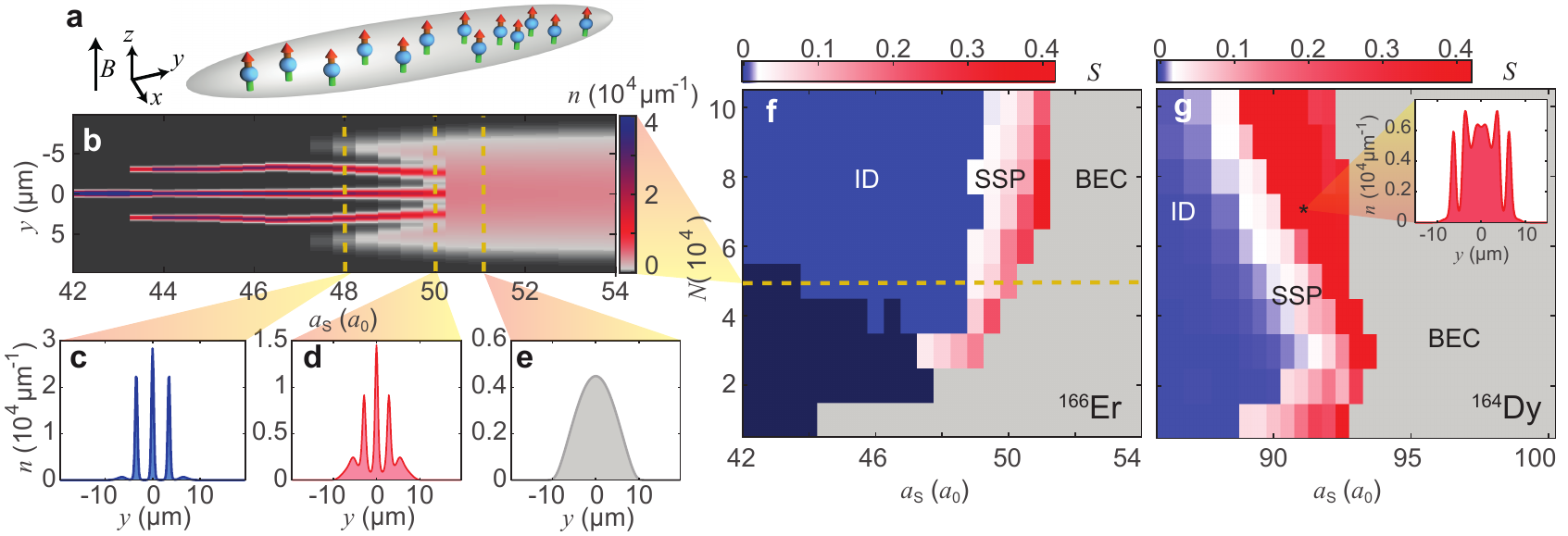}
	\caption{\label{fig:theory} \textbf{ 
	Phase diagram of an $\Er$ and a $\Dy$ dipolar BEC in a cigar-shaped trap.} (a) Illustration of the trap geometry with atomic dipoles oriented along $z$. (b) Integrated density profile as a function of $\as$ for an $\Er$ ground state of  $N=5\times 10^4$. In the colorbar, the density scale is upper-limited to $4\times 10^4\um^{-1}$ in order to enhance the visibility in the supersolid regime. (c--e) Exemplary density profiles for an insulating-droplet state (ID) at $\as=48\,a_0$, for a state with supersolid properties (SSP) at $50\,a_0$, and for a BEC at $51\,a_0$, respectively.  
	(f,\,g) Phase diagrams for $\Er$ and $\Dy$ for trap frequencies  $\omega_{x,y,z}=2\pi\times(145,31.5,151)\,$Hz and $2\pi\times(225,37,135)\,$Hz, respectively. The gray color identifies ground states with a single peak in $n(y)$ of large Gaussian width, $\sigma_y>2\ell_y$. The dark blue region in (f) shows the region where $n(y)$ exhibits a single sharp peak, $\sigma_y\leq2\ell_y$ and no density modulation. The red-to-blue colormap shows $S$ in the case of a density modulated $n(y)$. In (g) the colormap is upper-limited to 0.42 to enhance visibility in the low $N$ regime. The inset in (g) shows the calculated density profile for $\Dy$ at $N=7\times 10^4$ and $\as=91a_0$. 
	}
\end{figure*}

Droplet ground-states, quantum stabilization, and dipolar rotons have raised a huge excitement with very recent advancements adding key pieces of information to the supersolid scenario.  The quench experiments in an $\Er$ BEC at the roton instability have revealed out-of-equilibrium modulated states with an early-time phase coherence over a timescale shorter than a quarter of the oscillation period along the weak-trap axis \cite{Chomaz:2018}. In the same work, it has been suggested that the roton  softening combined with the quantum stabilization mechanism may open a promising route towards a supersolid ground state.  A first confirmation came from a recent theoretical work~\cite{Ancilotto:2018}, considering an Er BEC in an infinite elongated trap with periodic boundary conditions and tight transverse confinement. The supersolid phase appears to exist within a narrow region in interaction strength, separating a roton excitation with a vanishing energy and an incoherent assembly of insulating droplets.
Almost simultaneously, experiments with $\Dyo$ BECs in a shallow elongated trap, performing a slow tuning of the contact interaction, reported on the production of stripe states with phase coherence persisting up to half of the weak trapping period~\cite{Tanzi:2018}. More recently, such observations have been confirmed in another $\Dyo$ experiment~\cite{Bottcher2019}. Here, theoretical calculations showed the existence of a phase-coherent droplet ground-state, linking the experimental findings to the realization of a state with supersolid properties. The results on $\Dyo$ show however transient supersolid properties whose lifetime is limited by fast inelastic losses caused by three-body collisions~\cite{Tanzi:2018, Bottcher2019}.   These realizations raise the crucial question of whether a long-lived or stationary supersolid state can be created despite the usually non-negligble atom losses and the crossing of a discontinuous phase transition, which inherently creates excitations in the system.

In this work, we study both experimentally and theoretically 
the phase diagram of degenerate gases of highly magnetic atoms beyond the roton softening. Our investigations are carried out using two different experimental setups producing BECs of \Er~\cite{Aikawa:2012,Chomaz:2016} and of \Dy~\cite{Trautmann2018}, 
and rely on a fine tuning of the contact-interaction strength in both systems. 
In the regime of interest, these two atomic species have different contact-interaction scattering lengths, $\as$, whose precise dependence on the magnetic field is known only for Er~\cite{Baier2016ebh, Chomaz:2016, Chomaz:2018}, and different three-body-loss rate coefficients. Moreover, Er and Dy possess different magnetic moments, $\mu$, and masses, $m$, yielding the dipolar lengths, $\add=\mu_0\mu^2m/12\pi\hbar^2$, of $65.5\,a_0$ and $131\,a_0$, respectively. Here $\mu_0$ is the vacuum permeability, $\hbar=h/2\pi$ the reduced Planck constant, and $a_0$ the Bohr radius. For both systems, we find states showing hallmarks of supersolidity, namely the coexistence of density modulation and global phase coherence.   For such states, we quantify the extent of the $\as$-parameter range for their existence and study their lifetime.  For $\Er$, we find results very similar to the one recently reported for $\Dyo$~\cite{Tanzi:2018, Bottcher2019}, both systems being limited by strong three-body losses, which destroy the supersolid properties in about half of a trap period. However, for $\Dy$, we have identified an advantageous magnetic-field region where losses are very low and large BECs can be created. In this condition, we observe that the supersolid properties persist over a remarkably long time, well exceeding the trap period. Based on such a high stability, we finally demonstrate a novel route to reach the supersolid state, based on evaporative cooling from a thermal gas.

\section{Theoretical description}

As a first step in our study of the supersolid phase in dipolar BECs, we compute the  ground-state phase diagram for both $\Er$ and $\Dy$ quantum gases. The gases are confined in a cigar-shaped  harmonic trap, as illustrated in Fig.\,\ref{fig:theory}(a).  Our theory is based on numerical calculations of the extended Gross-Pitaevskii equation (eGPE) \cite{supmat}, which includes our anisotropic trapping potential, the short-range contact and long-range dipolar interactions at a mean-field level, as well as the first-order beyond-mean-field correction in the form of a Lee-Huang-Yang (LHY)  term~\cite{Waechtler:2016,Waechtler:2016b,Bisset:2016,Chomaz:2016,Chomaz:2018}. We note that, while both the exact strength of the LHY term and its dependence on the gas characteristics are under debate~\cite{Schmitt:2016,Chomaz:2018,Igor:2018, Cabrera2018, Petter:2018}, the importance of such a term, scaling with a higher power in density, is essential for stabilizing states beyond the mean-field instability~\cite{Schmitt:2016,Chomaz:2018,Igor:2018}; see also~\cite{Gammal:2000,Bulgac:2002,Lu2015sds,Petrov:2015}. 

Our theoretical results are summarized in Fig.\,\ref{fig:theory}. By varying the condensed-atom number, $N$, and $\as$, the phase diagram shows three very distinct phases. To illustrate them, we first describe the evolution of the integrated in-situ density profile $n(y)$ with fixed $N$ for varying $\as$, Fig.\,\ref{fig:theory}(b). The first phase, appearing at large $\as$, resembles a regular dilute BEC. It corresponds to a non-modulated density profile of low peak density and large axial size, $\sigma_y$, exceeding several times the corresponding harmonic oscillator length ($\ell_y=\sqrt{\hbar/m\omega_y}$), see Fig.\,\ref{fig:theory}(e) and the region denoted BEC in (f) and (g).  
The second phase appears when  decreasing $\as$ down to a certain critical value, $\as^*$. Here, the system undergoes an abrupt transition to a periodic density-modulated ground state, consisting of an array of overlapping narrow droplets, each of high peak density. Because the droplets are coupled to each other via a density overlap, later quantified in terms of the link strength $S$, particles can tunnel from one droplet to a neighboring one, establishing a global phase coherence across the cloud; see Fig.\,\ref{fig:theory}(d). Such a phase, in which periodic density modulation and phase coherence coexist, is identified as the supersolid one~\cite{Cinti2010sdc,Ancilotto:2018}; SSP region in (f) and (g). When further decreasing $\as$, we observe a fast reduction of the density overlap, which eventually vanishes; see Fig.\,\ref{fig:theory}(c). Here, the droplets become fully separated. Under realistic experimental conditions, it is expected that the phase relation between such droplets cannot be maintained; see later discussion. We identify this third phase as the one of an insulating droplet array~\cite{Bisset:2016,Waechtler1:2016b,Wenzel:2017}; ID region in (f) and (g). The number of droplets in the array decreases with  lowering either $\as$ (see (b)) or $N$, eventually resulting in a single droplet of high peak density, as in Refs.\,\cite{Waechtler:2016b,Bisset:2016}; see dark blue region in (f). The existence of these three phases (BEC-SSP-ID) is consistent with recent calculations considering an infinitely elongated Er BEC~\cite{Ancilotto:2018} and a cigar-shaped $\Dyo$ BEC~\cite{Bottcher2019}, illustrating the generality of this behavior in dipolar gases.

To study the supersolid character of the density-modulated phases,  
we compute the average of the wavefunction overlap between neighboring droplets, $S$.  As an ansatz to extract $S$, we use a Gaussian function to describe the wavefunction of each  individual droplet. This is found to be an appropriate description from an analysis of the density profiles of Fig.\,\ref{fig:theory}(b-d); see also \cite{Wenzel:2018}.  For two droplets at a distance $d$ and of identical Gaussian widths, $\sigma_y$, along the array direction, $S$ is simply $S=\exp(-d^2/4\sigma_y^2)$. Here, we generalize the computation of the wavefunction overlap to account for the difference in widths and amplitudes among neighboring droplets. 
This analysis allows to distinguish between the two types of modulated ground states, SSP and ID in Fig.\,\ref{fig:theory}(f--g). Within the Josephson-junction picture~\cite{Josephson:1962, Javanainen:1986,Javanainen:1986,Smerzi}, the tunneling rate of atoms between neighboring droplets depends on the wavefunction overlap, and an estimate for the single-particle tunneling rate can be derived within the Gaussian approximation \cite{Wenzel:2018}; see also~\cite{supmat}. The ID phase corresponds to vanishingly small values of $S$, yielding tunneling times extremely long compared to any other relevant time scale. In contrast, the supersolid phase is identified by a substantial value of $S$, with a correspondingly short tunneling time.

As shown in Fig.\,\ref{fig:theory}(f--g), a comparative analysis of the phase diagram for $\Er$ and $\Dy$ reveals similarities between the two species  (see also Ref.\,\cite{Bottcher2019}). A supersolid phase is found for sufficiently high $N$, in a narrow region of $\as$, upper-bounded by the critical value $\as^*(N)$. For intermediate $N$,  $\as^*$ increases with increasing $N$. We note that, for low $N$, the non-modulated BEC evolves directly into a single droplet state for decreasing $\as$ \cite{footnoteSD}. In this case, no supersolid phase is found in between, see also Refs.\,\cite{Waechtler:2016b,Bisset:2016}.
Despite the general similarities, we see that the supersolid phase for $\Dy$  appears for lower atom number than for Er and has a larger extension in $\as$. 
We note that, at large $N$ and for decreasing $\as$, Dy exhibits ground states with a density modulation appearing first in the wings, which then progresses inwards until a substantial modulation over the whole cloud is established \cite{footnotewings}; see inset (g). In this regime, we also observe that $\as^*$ decreases with increasing $N$. These type of states have not been previously reported and, although challenging to access in experiments because of the large $N$, they deserve further theoretical investigations.

\begin{figure*}[htbp!]
	\includegraphics[width=\textwidth]{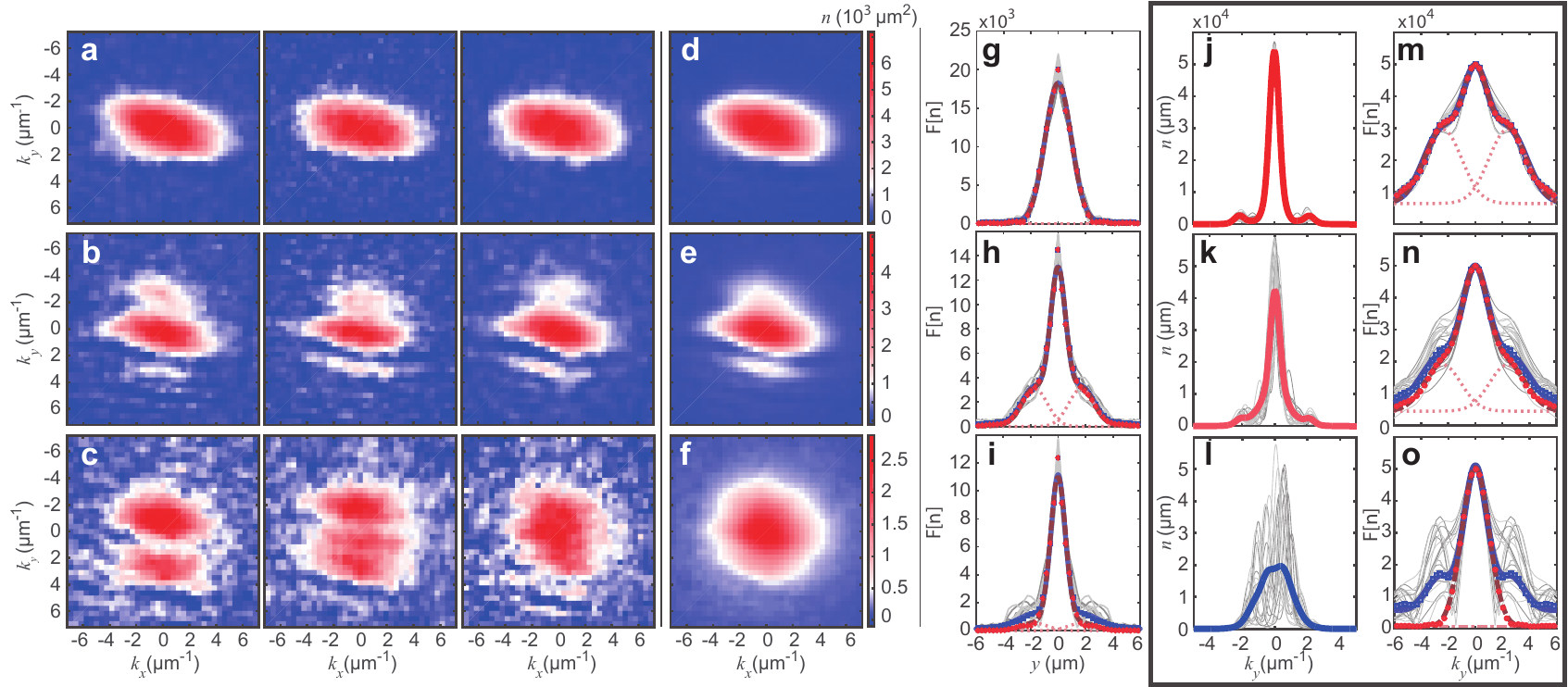}
	\caption{\label{fig:coherence} \textbf{Coherence in the interference patterns: measurement and toy model.} (a,b,c) Examples of single TOF absorption images at $\tho=5\,$ms for $\Er$ at $\as=\{54.7(2),53.8(2),53.3(2)\}\,a_0$, respectively. Corresponding average pictures for 100 images obtained under the same experimental conditions (d,\,e,\,f) and their FT profiles (g,\,h,\,i). The grey lines show the FT norm $|\mathcal{F}[n](y)|$ of the individual profiles. The averages, $n_\mathcal{M}$ (blue squares) and $n_\Phi$ (red dots), are fitted to three-Gaussian functions (blue solid line and brown dashed line, respectively). The dotted lines show the components of the total fitted function corresponding to the two side peaks in $n_\Phi$. (j-l) Interference patterns from the toy-model realizations with 100 independent draws using $N_D=4$,  $d=2.8\,\um$, $\sigma_y=0.56\,\um$ (see text) and for different $\phi_i$ distributions: (j) $\phi_i=0$, (k) $\phi_i$ normally distributed around 0 with $0.2\pi$ standard deviation, (l) $\phi_i$ uniformly distributed between 0 and $2\pi$. (m-o) Corresponding FT profiles for the toy model, same color code as (g-i). 
	}
\end{figure*}
\section{Experimental sequence for $^{166}$Erbium and $^{164}$Dysprosium}
To experimentally access the above-discussed physics, we produce dipolar BECs of either $\Er$ or $\Dy$ atoms. These two systems are created in different setups and below we summarize the main experimental steps; see also \cite{supmat}. 

\textit{Erbium} -- We prepare a stable $\Er$ BEC following the scheme of Ref.\,\cite{Chomaz:2018}.
At the end of the preparation, the Er BEC contains about $N=8\times 10^4$ atoms at $\as=64.5\,a_0$. The sample is  confined in a cigar-shaped optical dipole trap with harmonic frequencies $\omega_{x,y,z}=2\pi\times(145,31.5,151)\,$Hz. A homogeneous magnetic field, $B$, polarizes the sample along $z$ and controls the value of $\as$ via a magnetic Fesh\-bach resonance (FR)\,\cite{Chomaz:2016, Chomaz:2018,supmat}. Our measurements start by linearly ramping down $\as$  within $20\,$ms and waiting additional $15\,$ms so that $\as$ reaches its target value~\cite{supmat}. We note that ramping times between $20$ and $60\,$ms have been tested in the experiment and we do not record a significant difference in the system's behavior. 
After the 15\,ms stabilization time, we then hold the sample for a variable time $\tho$ before switching off the trap. Finally, we let the cloud expand for 30\,ms and perform absorption imaging along the $z$ (vertical) direction, from which we extract the density distribution of the cloud in momentum space, $n(k_x,k_y)$.

\textit{Dysprosium} -- The experimental procedure to create a $\Dy$\ BEC follows the one described in Ref.\,\cite{Trautmann2018}; see also \cite{supmat}. Similarly to Er, the Dy BEC is also confined in a cigar-shaped optical dipole trap and a homogeneous magnetic field ${B}$ sets the quantization axis along $z$ and the value of $\as$.
For Dy, we will discuss our results in terms of magnetic field, $B$, since the $\as$-to-$B$ conversion is not well known in the magnetic-field range considered~\cite{supmat,Tang2015,Schmitt:2016,Igor:2018}. 
In a first set of measurements, we first produce a stable BEC of about $N=3.5\times 10^4$ condensed atoms at a magnetic field of $B=2.5\,$G and then probe the phase diagram by tuning $\as$. Here, before ramping the magnetic field to access the interesting $\as$ regions, we slowly increase the power of the trapping beams within $200\,$ms. The final trap frequencies are  $\omega_{x,y,z}=2\pi\times(300,16,222)\,$Hz. 
After preparing a stable BEC, we ramp $B$ to the desired value within $20\,$ms and hold the sample for $\tho$~\cite{supmat}. In a second set of measurements, we study a completely different approach to reach the supersolid state. As discussed later, here we first prepare a thermal sample at a ${B}$ value where supersolid properties are observed and then further cool the sample until a transition to a coherent droplet-array state is reached.
In both cases, at the end of the experimental sequence, we perform absorption imaging after typically $27\,$ms of time-of-flight (TOF) expansion. The imaging beam propagates horizontally under an angle $\alpha$ of $\approx45^{\rm o}$ with respect to the weak axis of the trap ($y$). From the TOF images, we thus extract $n(k_Y,k_z)$ with $k_Y=\cos(\alpha)k_y+\sin(\alpha)k_x$. 

A special property of $\Dy$ is that its background scattering length is smaller than $\add$. This allows to enter the supersolid regime without the need of setting $B$ close to a FR, as done for $\Er$ and $\Dyo$, which typically causes severe atom losses due to increased three-body loss coefficients. In contrast, in the case of $\Dy$, the supersolid regime is reached by ramping $B$ away from the FR pole used to produce the stable BEC via evaporative cooling, as the $\as$-range of Fig.\,\ref{fig:theory}(g) lies  close to the background $\as$ reported in Ref.~\cite{Tang2015}; see also~\cite{supmat}. At the background level, three-body loss coefficients below $1.3\times 10^{-41} \text{m}^6 \text{s}^{-1}$ have been reported for $\Dy$~\cite{Schmitt:2016}.

\section{Density Modulation and phase coherence}

The coexistence of density modulation and phase coherence is the key feature that characterizes the supersolid phase and allows to discriminate it from the BEC and ID cases. 
To experimentally probe this aspect  in our dipolar quantum gases, we record their density distribution after a TOF expansion for various values of $\as$ across the phase diagram. As for a BEC in a weak optical lattice~\cite{Greiner2002} or for an array of BECs~\cite{Greiner2001,Paredes2004,Hadzibabic}, the appearance of interference patterns in the TOF images is associated with a density modulation of the in-situ atomic distribution. Moreover, the shot-to-shot reproducibility of the patterns (in amplitude and position) and the persistence of fringes in averaged pictures, obtained from many repeated images taken under the same experimental conditions, reveals the presence of phase coherence across the sample \cite{Hadzibabic}.

Figure~\ref{fig:coherence} exemplifies snapshots of the TOF distributions for Er, measured at three different $\as$ values; see (a-c). 
Even if very close in scattering length, the recorded $n(k_x,k_y)$ shows a dramatic change in behavior. For $\asf=54.7(2)\,a_0$, we observe a non-modulated distribution with a density profile characteristic of a dilute BEC. 
When lowering $\as$ to $53.8(2)\,a_0$, we observe the appearance of an interference pattern in the density distribution, consisting of a high central peak and two almost symmetric low-density side peaks\,\cite{Noteaberation}. 
Remarkably, the observed pattern is very reproducible with a high shot-to-shot stability, as shown in the repeated single snapshots and in the average image (b and e). This behavior indicates a coexistence of density modulation and global phase coherence in the in-situ state, as expected in the supersolid phase. This observation is consistent with our previous quench experiments~\cite{Chomaz:2018} and with the recent $\Dyo$ experiments~\cite{Tanzi:2018,Bottcher2019}. 
When further lowering $\asf$ to $53.3(2)\,a_0$, complicated patterns develop with fringes varying from shot-to-shot in number, position, and amplitude, signalizing the persistence of in-situ density modulation. However, the interference pattern is completely washed out in the averaged density profiles (f), pointing to the absence of a global phase coherence. We identify this behavior as the one of ID states.

\textit{Toy Model} -- To get an intuitive understanding of the interplay between density modulation and phase coherence and to estimate the role of the different sources of fluctuations in our experiment, we here develop a simple toy model, which is inspired by Ref.\,\cite{Hadzibabic}; see also Ref.~\cite{supmat}. 
In our model, the initial state is an array of $N_D$ droplets containing in total $N$ atoms. 
Each droplet is described by a one-dimensional Gaussian wavefunction, $\psi_i(y)$, of amplitude $\alpha_i$, phase $\phi_i$, width $\sigma_i$, and center $y_i$. 
To account for fluctuations in the experiments, we allow $\alpha_i$, $d_i=y_i-y_{i-1}$, and $\sigma_i$ to vary by 10\% around their expectation values. 
The spread of the phases ${\phi_i}$ among the droplets is treated specially as it controls the global phase coherence of the array. By fixing $\phi_i=0$ for each droplet or by setting a random distribution of $\phi_i$,  we range from full phase coherence to the incoherent cases. Therefore, the degree of phase incoherence can be varied by changing the standard deviation of  the distribution of $\phi_i$.  

To mimic our experiment, we  compute the free evolution of each individual $\psi_i$  over $30\,$ms, and then compute the axial distribution $n(y,t) = |\sum_i \psi_i(y,t)|^2$,  
from which we extract the momentum distribution $n(k_y)$, also accounting for the finite imaging resolution~\cite{supmat}. 
For each computation run, we randomly draw $N_D$ values for $\phi_i$, as well as of $\sigma_i$, $d_i$ and $\alpha_i$ 
and extract $n(k_y)$. We then collect a set of $n(k_y)$ by drawing these values multiple times using the same statistical parameters and compute the expectation value, $ \langle n(k_y) \rangle$; see Fig.\,\ref{fig:coherence}(j-l). The angled brackets denote the ensemble average.

The results of our toy model show large similarity with the observed behavior in the experiment.
In particular, while for each single realization one can clearly distinguish multi-peak structures regardless of the degree of phase coherence between the droplets, the visibility of the interference pattern in the averaged $n(k_y)$ survives only if the standard deviation of the phase fluctuations between droplets is small (roughly, below $0.3\pi$). In the incoherent case, we note that the shape of the patterns strongly vary from shot to shot. Interestingly, the toy model also shows that the visibility of the coherent peaks in the average images is robust against the typical shot-to-shot fluctuations in droplet size, amplitude and distance that occur in the experiments; see (j,k).

\textit{Probing density modulation and phase coherence} -- 
To separate and quantify the information on the in-situ density modulation and its phase coherence, we analyse the measured interference patterns in Fourier space ~\cite{takeda1982ftm,Kohstall2011,Chomaz:2015eoc,Bottcher2019}. Here, we extract two distinct averaged density profiles, $n_{\mathcal{M}}$ and $n_\Phi$. Their structures at finite $y$-spatial frequency (i.\,e.\,in Fourier space) quantify the two above-mentioned properties.

More precisely, we perform a Fourier transform (FT) of the integrated momentum distributions, $n(k_y)$, denoted $\mathcal{F}[n](y)$. 
Generally speaking, modulations in $n(k_y)$ induce peaks at finite spatial frequency, $y=y^*$, in the FT norm, $\left|\mathcal{F}[n](y)\right|$; see Fig.\,\ref{fig:coherence}(g-i) and (m-o). Following the above discussion (see also Refs.~\cite{Hadzibabic,hofferberth2007nec}), such peaks in an individual realization hence reveal a density modulation of the corresponding in-situ state, with a wavelength roughly equal to $y^*$. Consequently, we consider the  average of the FT norm of the individual images, $n_\mathcal{M}(y)= \langle |\mathcal{F}[n](y)|\rangle$ as the first profile of interest. 
The peaks of $n_\mathcal{M}$ at finite $y$ then indicate the  mere existence of an in-situ density modulation of roughly constant spacing within the different realizations. As second profile of interest, we use the FT norm of the average profile $\langle n(k_y) \rangle$, $n_\Phi(y)=|\mathcal{F}[\langle n\rangle](y)|$.
Connecting to our previous discussion,  
the peaks of $n_\Phi$ at finite $y$ point to the persistence of a modulation in the average $\langle n(k_y) \rangle$, which we identified as a hallmark for a global phase coherence within the density modulated state. In particular we point out that a perfect phase coherence, implying identical interference patterns in all the individual realizations, yields $n_\mathcal{M}=n_\Phi$ and, thus,  identical peaks at finite $y$ in both profiles. We note that, by linearity, $n_\Phi$, also matches the norm of the average of the full FT of the individual images, i.\,e.\,$n_\Phi(y)=|\langle \mathcal{F}[n](y)\rangle|$; see also Ref.~\cite{supmat}.

Figure\,\ref{fig:coherence} (g-i) and (m-o) demonstrates the significance of our FT analysis scheme by applying it to the momentum distributions from the experiment (d-f) and the ones from the toy model (j-l), respectively. 
As expected, for the BEC case, both $n_\mathcal{M}$ and $n_\Phi$ show a single peak at zero spatial frequency, $y=0$,  characterizing the absence of density modulation (g). 
In the case of phase-coherent droplets (e), we observe that $n_\mathcal{M}$ and $n_\Phi$ are superimposed and both show two symmetric side peaks at finite $y$, 
in addition to a dominant peak at $y=0$;  
see (h). In the incoherent droplet case, we find that, while  $n_\mathcal{M}$ still shows side peaks at finite $y$, 
the ones in  $n_\Phi$  wash out from the averaging, (f,\,i,\,l,\,o). For both coherent and incoherent droplet arrays, the toy-model results show behaviors matching the above description, providing a further justification of our FT-analysis scheme; see (j-o). Our toy model additionally proves two interesting features. First, it shows that the equality $n_\mathcal{M}=n_\Phi$, revealing the global phase coherence of a density modulated state, is remarkably robust to noise in the structure of the droplet arrays; see (j,\,m). Second, our toy model however shows that phase fluctuations across the droplet array on the order of 0.2$\pi$ standard deviation are already sufficient to make $n_\Phi$ and $n_\mathcal{M}$ to deviate from each other; see (k,\,n). The incoherent behavior is also associated with strong variations in the side peak amplitude of the individual realizations of $|\mathcal{F}[n]|$, connecting, e.\,g.\,to the observations of Refs.\,\cite{Bottcher2019}.  

Finally, to quantify the density modulation and the phase coherence, we fit a three-Gaussian function to both $n_\mathcal{M}(y)$ and $n_\Phi(y)$ and extract the amplitudes of the finite-spatial-frequency peaks, $A_{\mathcal{M}}$ and $A_{\Phi}$, for both distributions, respectively. Note that for a BEC, which is a phase coherent state, $A_{\Phi}$ will be zero since it probes only finite-spatial-frequency peaks; see Fig.\,\ref{fig:coherence}\,(g-i,m-o).

\begin{figure}[t!]
	\includegraphics[width=1\linewidth]{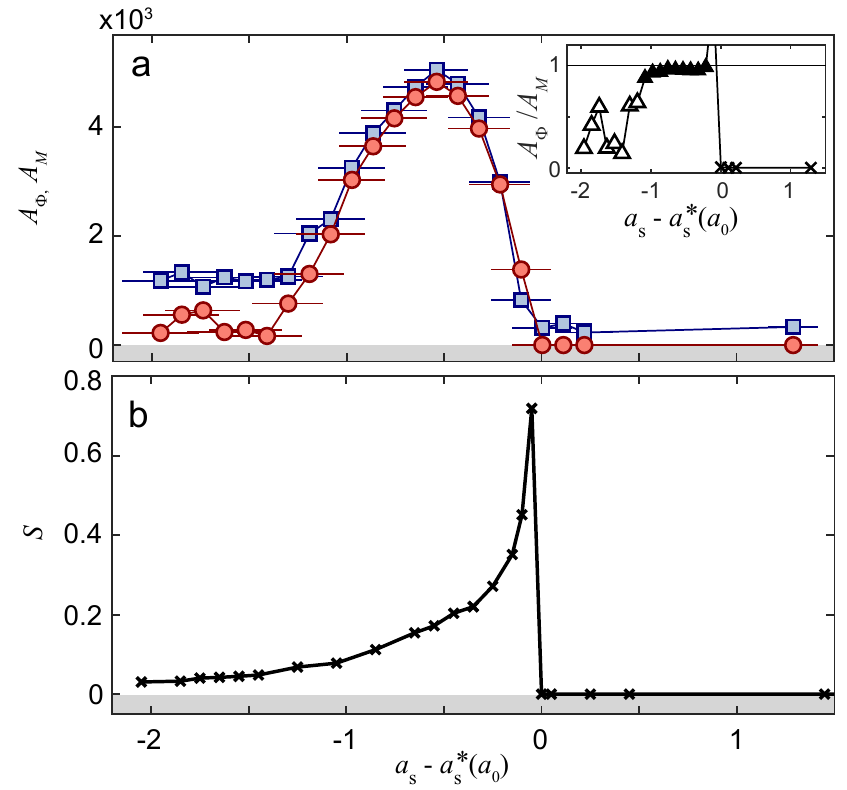}
	\caption{\label{fig:ContrastVSa} \textbf{Supersolid behavior across the phase diagram.} Measured side-peak amplitudes, $A_{\Phi}$ (circles) and $A_{\mathcal{M}}$ (squares)  with their ratio in inset (a), and calculated link strength $S$ (b) as a function of $\as-a_s^{*}$ for \Er. For non-modulated states, we set $S=0$ in theory and $A_{\Phi}/A_{\mathcal{M}}=0$ in experiment (crosses in inset). In the inset, open and closed symbols correspond to $A_{\Phi}/A_{\mathcal{M}}>0.8$ and $\leq0.8$, respectively. In the experiments, 
	we probe the system at a fixed $\tho=5\,$ms. Horizontal error bars are derived from our experimental uncertainty in $B$, vertical error bars corresponding to the statistical uncertainty from the fit are smaller than the data points.   The measured and calculated critical scattering lengths are $a_s^{*}=54.9(2)\,a_0$   and $50.55\,a_0$, respectively~\cite{footnoteastarvalue}.
	The numerical results are obtained for the experimental trap frequencies and for a constant $N=5\times10^4$~\cite{NoteN}.}
\end{figure}

\section{Characterization of the Supersolid State} 
We are now in the position to study two key aspects, namely (i) the evolution of the density modulation and phase coherence across the BEC-supersolid-ID phases, and (ii) the lifetime of the coherent density modulated state in the supersolid regime.

{\em Evolution of the supersolid properties across the phase diagram} -- The first type of investigations is conducted with $\Er$ since for this species the scattering length and its dependence on the magnetic field has been precisely characterized\,\cite{Chomaz:2016,Chomaz:2018}. After preparing the sample, we ramp $\as$ to the desired value and  study the density  patterns as well as their phase coherence by probing the amplitudes $A_{\mathcal{M}}$ and $A_{\Phi}$ as a function of $\as$ after $\tho=5\,$ms. 
As shown in Fig.\,\ref{fig:ContrastVSa}(a), in the BEC region (i.\,e.\,for large $\as$), we observe that both $A_{\mathcal{M}}$ and $A_\Phi$  are almost zero, evidencing the expected absence of a density modulation in the system. As soon as $\as$ reaches a critical value $\as^*$,  the system's behavior dramatically changes with a sharp and simultaneous increase of both $A_{\mathcal{M}}$ and $A_\Phi$. 
While the strength of $A_{\mathcal{M}}$ and $A_\Phi$ varies with decreasing $\as$ -- first increasing then decreasing -- we observe that their ratio $A_\Phi/A_{\mathcal{M}}$ remains constant and close to unity over a narrow $\as$-range below $\as^*$ of $\gtrsim 1\,a_0$ width; see inset. This behavior pinpoints the coexistence in the system of phase coherence and density modulation, as predicted to occur in the supersolid regime.
For $(\as-\as^*)< -1\,a_0$, we observe that the two amplitudes depart from each other. Here, while the density modulation still survives
with $A_{\mathcal{M}}$ saturating to a lower finite value, the global phase coherence is lost with $A_{{\Phi}}/A_{\mathcal{M}}<1$, 
as expected in the insulating droplet phase.

To get a deeper insight on how our observations compare to the phase-diagram predictions (see Fig.\,\ref{fig:theory}), we study the link strength $S$ as a function of $\as$; see Fig.\,\ref{fig:ContrastVSa}(b).
Since $S$ quantifies the density overlap between neighboring droplets and is related to the tunneling rate of atoms across the droplet array, it thus provides information on the ability of the system to establish or maintain a global phase coherence. In this plot, we set $S=0$ in the case where no modulation is found in the ground state. 
At the BEC-to-supersolid transition, i.\,e.\,at $\as=\as^*$, a density modulation abruptly  
appears in the system's ground-state with  $S$ taking a finite value. 
Here, $S$ is maximal, corresponding to a density modulation of minimal amplitude. Below the transition, we observe a progressive decrease of $S$ with lowering $\as$, pointing to the gradual reduction of the tunneling rate in the droplet arrays. Close to the transition, we estimate a large tunneling compared to all other relevant timescales. However, we expect this rate to become vanishingly small, on the sub-Hertz level~\cite{supmat}, when decreasing $\as$ $1$--$2\,a_0$ below $\as^*$.  Our observation also hints to the smooth character of the transition from a supersolid to an ID phase, suggesting a crossover-type behavior.

The general trend of $S$, including the extension in $\as$ where it takes non-vanishing values, is similar to the $\as$-behavior of $A_{\mathcal{M}}$ and $A_{\Phi}$ observed in the experiments ~\cite{footnoteastarvalue}. We observe in the experiments that the $\as$ dependence at the BEC-to-supersolid transition appears sharper than at the supersolid-to-ID interface, potentially suggesting a different nature of the two transitions. However, more investigations are needed since atom losses, finite-temperature and finite-size effects can affect, and in particular smoothen, the  observed behavior~\cite{Fisher1967,Imry:1971,Imry:1980}. Moreover, dynamical effects, induced by e.\,g.\, excitations created at the crossing of the phase transitions or atoms losses during the time evolution, can also play a substantial role in the experimental observations, complicating a direct comparison with the ground state calculations. The time dynamics as well as a different scheme to achieve a state with supersolid properties will be the focus of the remainder of the manuscript. 

\begin{figure}[t!]
	\includegraphics{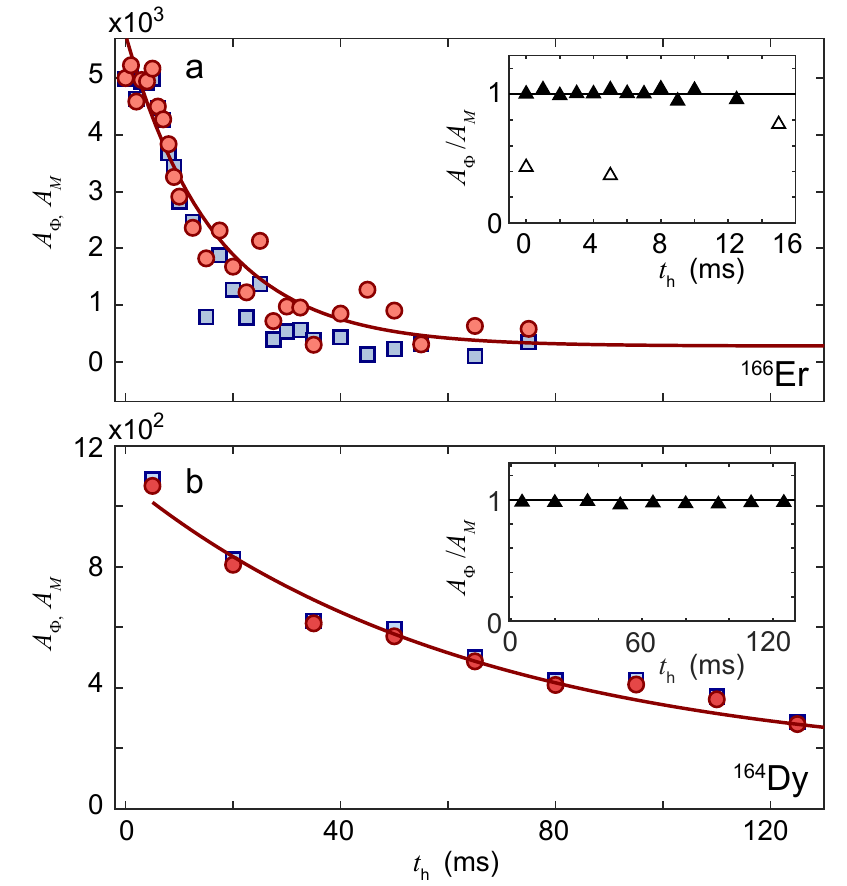}
	\caption{\label{fig:DecayRateCoherenceA} 
	\textbf{Time Evolution of the supersolid properties.} Amplitudes $A_\Phi$ (circles) and $A_{\mathcal{M}}$ (squares) in the supersolid regime as a function of the holding time in trap for (a) $\Er$ at $54.2(2)\,a_0$, and for (b) $\Dy$ at 2.04\,G. 
	The solid lines are exponential fits to the data. The insets show the time evolution of $A_\Phi/A_{\mathcal{M}}$ for the above cases (filled triangles), and, for comparison, in the ID regime (empty triangles) for Er at $\as = 53.1(2)\,a_0$ (a).} 
\end{figure}
{\em Lifetime of the supersolid properties} -- Having identified the $\as$ range in which our dipolar quantum gas exhibits supersolid properties, the next central question concerns the stability and lifetime of such a fascinating state. Recent experiments on $\Dyo$ have shown the transient character of the supersolid properties, whose lifetime is limited by three-body losses~\cite{Tanzi:2018,Bottcher2019}. In these experiments, the phase coherence is found to survive up to 20\,ms after the density modulation has formed. This time corresponds to about half of the weak-trap period. Stability is a key issue in the supersolid regime, especially since the tuning of $\as$, used to enter this regime, has a twofold consequence on the inelastic loss rate. First, it gives rise to an increase in the peak density (see Fig.\,\ref{fig:theory}b-d) and, second, it may lead to an enhancement of the three-body loss coefficient. 

We address this question by conducting comparative studies on $\Er$ and $\Dy$ gases. 
These two species allow us to tackle two substantially different scattering scenarios. Indeed, the background value of $\as$ for $\Er$ (as well as for $\Dyo$) is larger than $\add$. Thus, reaching the supersolid regime, which occurs at $\add /\as \approx 1.2-1.4$ in our geometry, requires to tune ${B}$ close to the pole of a FR. 
This tuning also causes an increase of the three-body loss rate.  In contrast, $\Dy$ realizes the opposite case with the background scattering length smaller than $\add$. This feature brings the important advantage of requiring tuning $B$ away from the FR pole to reach the supersolid regime. As we will describe below, this important difference in scattering properties leads to a strikingly longer lifetime of the $\Dy$ supersolid properties with respect to \Er and to the recently observed behavior in \Dyo~\cite{Tanzi:2018,Bottcher2019}.

The measurements proceed as follows. For both $\Er$ and $\Dy$, we first prepare the quantum gas in the stable BEC regime and then ramp $\as$ to a fixed value in the supersolid regime for which the system exhibits a state of coherent droplets (i.\,e.\,$A_\Phi / A_{\mathcal{M}} \approx 1$); see previous discussion.
Finally, we record the TOF images after a variable $\tho$ and we extract the time evolution of both $A_\Phi$ and $A_{\mathcal{M}}$. The study of these two amplitudes will allow us to answer the question whether the droplet structure -- i.\,e.\,the density modulation in space -- persists in time whereas the coherence among droplets is lost ($ A_{\mathcal{M}} >  A_\Phi \rightarrow 0$) or if the density structures themselves vanish in time ($ A_{\mathcal{M}} \approx A_\Phi \rightarrow 0$).  
\begin{figure}[t!]
	\includegraphics[width=1\linewidth]{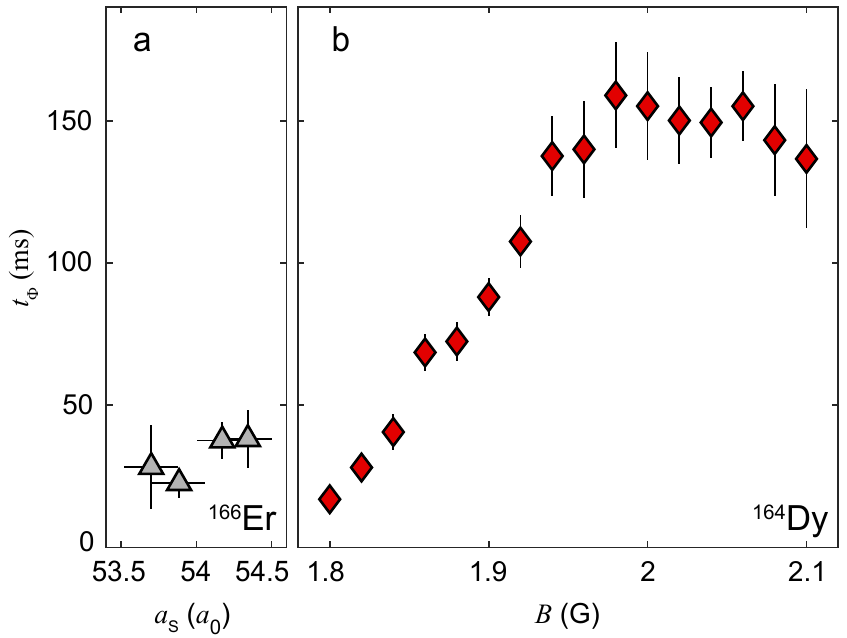}
	\caption{\label{fig:DecayRateCoherence} 
 \textbf{Survival time of the coherent density-modulated state.}  $t_\Phi$ in $\Er$ as a function of $\as$ (a) and $\Dy$ as a function of $B$ (b). The error bars refer to the statistical uncertainty from the fit. The range of investigation corresponds to the supersolid regime for which phase-coherent density-modulated states are observed. This range is particularly narrow for $\Er$.}
\end{figure}

As shown in Fig.\,\ref{fig:DecayRateCoherenceA}, for both species, we observe that  $A_\Phi$ and $A_{\mathcal{M}}$ decay almost synchronously with a remarkably longer lifetime for $\Dy$ (b) than $\Er$ (a). Interestingly, $A_\Phi$ and $A_\mathcal{M}$ remain approximately equal during the whole time dynamics; see inset (a-b). This behavior indicates that it is the strength of the density modulation itself and not the phase coherence among droplets that decays over time. Similar results have been found theoretically in Ref.\,\cite{footnoteLuis}. We connect this decay mainly to three-body losses, especially detrimental for $\Er$, and possible excitations created while crossing the BEC-to-supersolid phase transition~\cite{supmat}. For comparison, the inset shows also the behavior in the ID regime for $\Er$, where $A_\Phi/A_\mathcal{M}<1$ already at 
short $\tho$ and remains so during the time evolution~\cite{supmat}.

To get a quantitative estimate of the survival time of the phase coherent and density modulated state, we fit a simple exponential function to $A_\Phi$ and extract $t_\Phi$, defined as the $1/10$ lifetime; see Fig.\,\ref{fig:DecayRateCoherenceA}. For $\Er$, we extract $t_\Phi =38(6)\,$ms. For $\tho>t_\Phi$, the interference patterns become undetectable in our experiment and we recover a signal similar to the one of a non-modulated BEC state (as in Fig.\,\ref{fig:coherence}(a,d)). These results are consistent with recent observations of transient supersolid properties in $\Dyo$~\cite{Tanzi:2018}. For $\Dy$, we observe that the coherent density-modulated state is remarkably long-lived. Here, we find $t_\Phi=152(13)\,$ms. 

The striking difference in the lifetime and robustness of the supersolid properties between $\Er$ and $\Dy$ becomes even more visible when studying $t_\Phi$ as a function of $\as$ (${B}$ for Dy). As shown in  Fig.\,\ref{fig:DecayRateCoherence}, $t_\Phi$ for Er remains comparatively low in the investigated supersolid regime and slightly varies between 20 and 40\,ms. Similarly to the recent studies with $\Dyo$, this finding reveals the transient character of the state and opens the question of whether a stationary supersolid state can be reached with these species. On the contrary, for $\Dy$ we observe that $t_\Phi$ first increases with ${B}$ in the range from $1.8\,$G to about $1.98\,$G. Then, for ${B}>1.98\,$G, $t_\Phi$ acquires a remarkably large and  almost constant value of about $150\,$ms over a wide ${B}$-range. This shows the long-lived character of the supersolid properties in our $\Dy$ quantum gas. We note that over the investigated range, $\as$ is expected to monotonously increase with ${B}$~\cite{supmat}. Such a large value of $t_\Phi$ exceeds not only the estimated tunneling time across neighbouring droplets but also the weak-axis trap period, which together set the typical timescale to achieve global equilibrium and to study collective excitations.

\begin{figure}[htbp!]
	\includegraphics[width=1\linewidth]{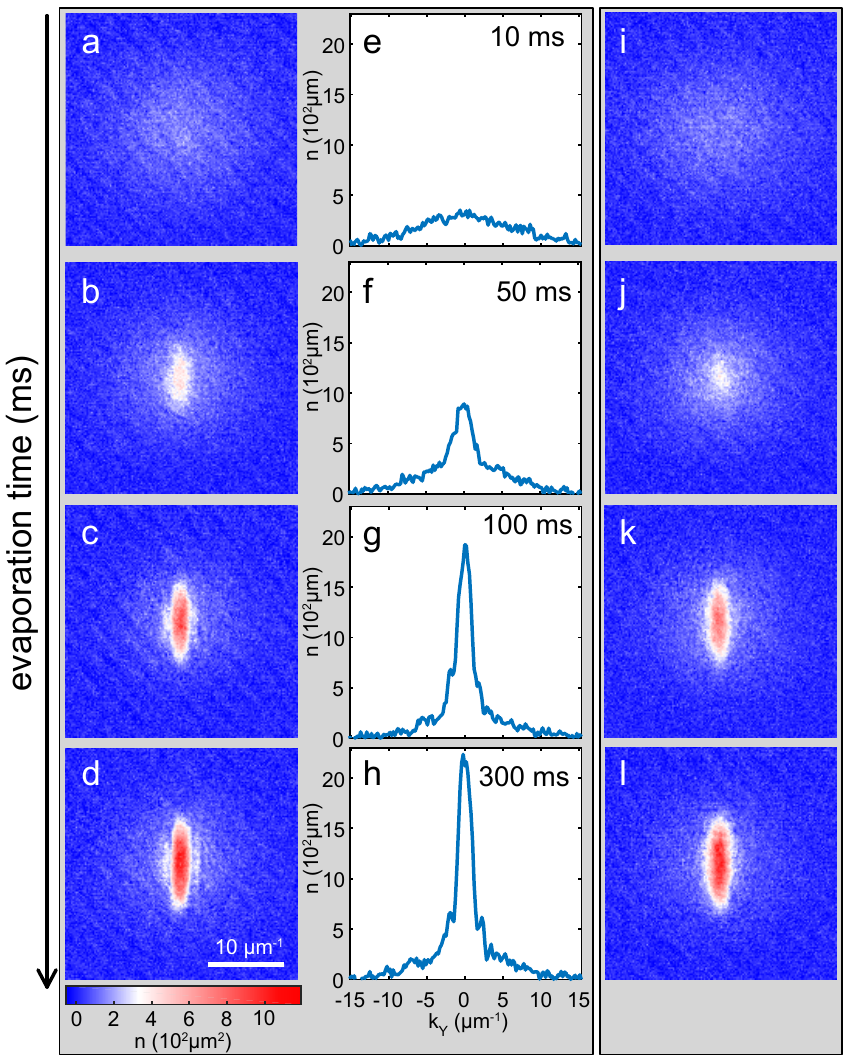}
	\caption{\label{fig:Evap}  \textbf{Evaporative cooling to a state with supersolid properties.} $\Dy$ absorption images showing the transition to a state with supersolid properties at 2.43\,G (a-d) and to a BEC state at 2.55\,G (i-l), via different durations of the last evaporation step. These durations are $10\,$ms (a,i), $50\,$ms (b,j), $100\,$ms (c,k), and $300\,$ms (d,l).  The density profiles (e-h) are integrated over the central regions of the corresponding absorption images (a-d). The colormap indicates the atomic density in momentum space.
	}
\end{figure}

\section{Creation of states with supersolid properties by evaporative cooling} 

The long-lived supersolid properties in $\Dy$ motivate us to explore an alternative route to cross the supersolid phase transition, namely by evaporative cooling instead of interaction tuning. 
For this set of experiments, we have modified the waists of our trapping beams in order to achieve quantum degeneracy in tighter traps with respect to the one used for condensation in the previous set of measurements. In this way, the interference peaks in the supersolid region are already visible without the need to apply a further compression of the trap since the side-to-central-peak distance in the momentum distribution scales roughly as $1/\ell_z$ \cite{Chomaz:2018}. Forced evaporative cooling is performed by  reducing the power of the trapping beams piecewise-linearly in subsequent evaporation steps until a final trap with frequencies $2\pi\times(225,37,134)\,$Hz is achieved. During the whole evaporation process, which has an overall duration of about 3\,s, the magnetic field is kept either at ${B}=2.43\,$G, where we observe long-lived interference patterns, or at ${B}=2.55\,$G, where we produce a stable non-modulated BEC. We note that these two $B$ values are very close without any FR lying in between~\cite{supmat}. 

Figure \ref{fig:Evap} shows the phase transition from a thermal cloud to a final state with supersolid properties by evaporative cooling. 
In particular, we study the phase transition by varying the duration of the last evaporation ramp, while maintaining the initial and final trap-beam power fixed. This procedure effectively changes the atom number and temperature in the final trap while keeping the trap parameters unchanged, which is important to not alter the final ground-state phase diagram of the system. At the end of the evaporation, we let the system equilibrate and thermalize for $\tho = 100\,$ms, after which we switch off the trap, let the atoms expand for $26.5\,$ms, and finally perform absorption imaging. We record the TOF images for different ramp durations, i.\,e.\ for different thermalization times. For a short ramp,  too many atoms are lost such that the critical atom number for condensation is not reached, and the atomic distribution remains thermal; see Fig.\,\ref{fig:Evap}(a). 

By increasing the ramp time, the evaporative cooling becomes more efficient and we observe the appearance of a bimodal density profile with a narrow and dense peak at the center, which we identify as a regular BEC; see Fig.\,\ref{fig:Evap}(b). By further cooling, the BEC fraction increases and the characteristic pattern of the supersolid state emerges; see Fig.\,\ref{fig:Evap}(c-d). The observed evaporation process shows a strikingly different behavior in comparison with the corresponding situation at $B=2.55\,$G, where the usual thermal-to-BEC phase transition is observed; see Fig.\,\ref{fig:Evap}(i-l). 

\begin{figure}[htbp!]
	\includegraphics[width=1\linewidth]{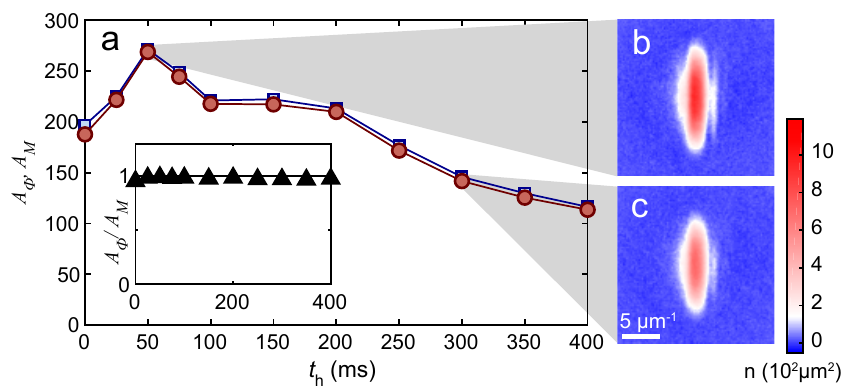}
	\caption{\label{fig:EvapLife} \textbf{Lifetime of the supersolid properties achieved via evaporative cooling.} Time evolution of the amplitudes $A_\Phi$ (red circle) and $A_{\mathcal{M}}$ (square) after an evaporation time of $300\,$ms at 2.43\,G and an equilibration time of $100\,$ms. The inset shows the time evolution of $A_\Phi/A_{\mathcal{M}}$. Averaged absorption images of 25 realizations after $50\,$ms (b) and $300\,$ms (c) of holding time. Note that the thermal background has been substracted from the images. The colormap indicates the atomic density in momentum space.
	}
\end{figure}

We finally probe the lifetime of the supersolid properties by extracting the time evolution of both the amplitudes $A_\Phi$ and $A_{\mathcal{M}}$, as previously discussed. We use the same experimental sequence as the one in Fig.\,\ref{fig:Evap}(d) -- i.\,e.\,$300\,$ms duration of the last evaporation ramp and $100\,$ms of equilibration time -- and subsequently hold the sample in the trap for a variable $\tho$. As shown in Fig.\,\ref{fig:EvapLife}(a), we observe a very long lifetime with both amplitudes staying large and almost constant over more than 200\,ms. At longer holding time, we observe a slow decay of $A_\Phi$ and $A_{\mathcal{M}}$, following the one of the atom number.  Moreover, during the dynamics, the ratio $A_\Phi/A_{\mathcal{M}}$ stays constant. The long lifetime of the phase-coherent density modulation is also directly visible in the persistence of the interference patterns in the averaged momentum density profiles (similar to Fig.\,\ref{fig:coherence}(e)), both at intermediate and long times; see Fig.\ref{fig:EvapLife}(b) and (c), respectively.  For even longer $\tho$, we can not resolve anymore interference patterns in the TOF images. Here, we recover a signal consistent with a regular BEC of low $N$.  

Achieving the coherent-droplet phase via evaporative cooling is a very powerful alternative path to supersolidity. We speculate that, for instance, excitations, which might be important when crossing the phase transitions by interaction tuning, may be small or removed by evaporation when reaching this state kinematically. Other interesting questions, open to future investigations, are the nature of the phase transition, the critical atom number, and the role of non-condensed atoms.

\section{Conclusions} 
For both $\Er$ and $\Dy$ dipolar quantum gases, we have identified and studied states showing hallmarks of supersolidity, namely global phase coherence and spontaneous density modulations. These states exist in a narrow scattering-length region, lying between a regular BEC phase and a phase of an insulating droplet array. 
While for $\Er$, similarly to the recently-reported $\Dyo$ case \cite{Tanzi:2018, Bottcher2019}, the observed supersolid properties fade out over a comparatively short time because of atom losses, we find that $\Dy$ exhibits remarkably long-lived supersolid properties. 
Moreover, we are able to directly create stationary states with supersolid properties by evaporative cooling, demonstrating a powerful alternative approach to interaction tuning on a BEC. This novel technique provides prospects of creating states with supersolid properties while avoiding 
additional excitations and dynamics. 
The ability to produce long-lived supersolid states paves the way for future investigations on quantum fluctuations and many-body correlations as well as of collective excitations in such an intriguing many-body quantum state. 
A central goal of these future investigations lies in proving the superfluid character of this phase, beyond its global phase coherence~\cite{Boninsegni:2012,Scarola2006,Lu2015,Cinti2017,Ancilotto:2018}.

\textit{Note added}-- During the course of the present work, we became aware of a related work, reporting a theoretical study of the ground-state phase diagram based on  Monte-Carlo calculations \cite{Youssef2019}.

\section{Acknowledgments} 
\begin{acknowledgments}
We thank  R.\,Bisset, B.\,Blakie, M.\,Boninsegni, G.\,Modugno, T.\,Pfau, and in particular L.\,Santos for many stimulating discussions.
Part of the computational results presented have been achieved using the HPC infrastructure LEO of the University of Innsbruck. 
We acknowledge support by the Austrian Science Fund FWF through the DFG/FWF Forschergruppe (FOR 2247/PI2790), by the ERC Consolidator Grant (RARE, no.~681432), and by a NFRI Grant (MIRARE, no.~\"OAW0600) from the Austrian Academy of Science. G.\,D.~and M.\,S.~acknowledge support by the Austrian Science Fund FWF within the DK-ALM (no.~W1259-N27). 
\end{acknowledgments}

* Correspondence and requests for materials
should be addressed to Francesca.Ferlaino@uibk.ac.at.

\bibliography{SupersolidBib}

\begin{thebibliography}{78}%
\makeatletter
\providecommand \@ifxundefined [1]{%
 \@ifx{#1\undefined}
}%
\providecommand \@ifnum [1]{%
 \ifnum #1\expandafter \@firstoftwo
 \else \expandafter \@secondoftwo
 \fi
}%
\providecommand \@ifx [1]{%
 \ifx #1\expandafter \@firstoftwo
 \else \expandafter \@secondoftwo
 \fi
}%
\providecommand \natexlab [1]{#1}%
\providecommand \enquote  [1]{``#1''}%
\providecommand \bibnamefont  [1]{#1}%
\providecommand \bibfnamefont [1]{#1}%
\providecommand \citenamefont [1]{#1}%
\providecommand \href@noop [0]{\@secondoftwo}%
\providecommand \href [0]{\begingroup \@sanitize@url \@href}%
\providecommand \@href[1]{\@@startlink{#1}\@@href}%
\providecommand \@@href[1]{\endgroup#1\@@endlink}%
\providecommand \@sanitize@url [0]{\catcode `\\12\catcode `\$12\catcode
  `\&12\catcode `\#12\catcode `\^12\catcode `\_12\catcode `\%12\relax}%
\providecommand \@@startlink[1]{}%
\providecommand \@@endlink[0]{}%
\providecommand \url  [0]{\begingroup\@sanitize@url \@url }%
\providecommand \@url [1]{\endgroup\@href {#1}{\urlprefix }}%
\providecommand \urlprefix  [0]{URL }%
\providecommand \Eprint [0]{\href }%
\providecommand \doibase [0]{http://dx.doi.org/}%
\providecommand \selectlanguage [0]{\@gobble}%
\providecommand \bibinfo  [0]{\@secondoftwo}%
\providecommand \bibfield  [0]{\@secondoftwo}%
\providecommand \translation [1]{[#1]}%
\providecommand \BibitemOpen [0]{}%
\providecommand \bibitemStop [0]{}%
\providecommand \bibitemNoStop [0]{.\EOS\space}%
\providecommand \EOS [0]{\spacefactor3000\relax}%
\providecommand \BibitemShut  [1]{\csname bibitem#1\endcsname}%
\let\auto@bib@innerbib\@empty
\bibitem [{\citenamefont {Andreev}\ and\ \citenamefont
  {Lifshitz}(1969)}]{Andreev}%
  \BibitemOpen
  \bibfield  {author} {\bibinfo {author} {\bibfnamefont {A.~F.}\ \bibnamefont
  {Andreev}}\ and\ \bibinfo {author} {\bibfnamefont {I.~M.}\ \bibnamefont
  {Lifshitz}},\ }\href@noop {} {\bibfield  {journal} {\bibinfo  {journal} {Sov.
  Phys. JETP}\ }\textbf {\bibinfo {volume} {29}},\ \bibinfo {pages} {1107}
  (\bibinfo {year} {1969})}\BibitemShut {NoStop}%
\bibitem [{\citenamefont {Chester}(1970)}]{Chester}%
  \BibitemOpen
  \bibfield  {author} {\bibinfo {author} {\bibfnamefont {G.~V.}\ \bibnamefont
  {Chester}},\ }\href {\doibase 10.1103/PhysRevA.2.256} {\bibfield  {journal}
  {\bibinfo  {journal} {Phys. Rev. A}\ }\textbf {\bibinfo {volume} {2}},\
  \bibinfo {pages} {256} (\bibinfo {year} {1970})}\BibitemShut {NoStop}%
\bibitem [{\citenamefont {Leggett}(1970)}]{Leggett}%
  \BibitemOpen
  \bibfield  {author} {\bibinfo {author} {\bibfnamefont {A.~J.}\ \bibnamefont
  {Leggett}},\ }\href {\doibase 10.1103/PhysRevLett.25.1543} {\bibfield
  {journal} {\bibinfo  {journal} {Phys. Rev. Lett.}\ }\textbf {\bibinfo
  {volume} {25}},\ \bibinfo {pages} {1543} (\bibinfo {year}
  {1970})}\BibitemShut {NoStop}%
\bibitem [{\citenamefont {Kirzhnits}\ and\ \citenamefont
  {Nepomnyashchii}(1971)}]{Kirzhnits:1971cco}%
  \BibitemOpen
  \bibfield  {author} {\bibinfo {author} {\bibfnamefont {D.}~\bibnamefont
  {Kirzhnits}}\ and\ \bibinfo {author} {\bibfnamefont {Y.~A.}\ \bibnamefont
  {Nepomnyashchii}},\ }\href@noop {} {\bibfield  {journal} {\bibinfo  {journal}
  {Sov. Phys. JETP}\ }\textbf {\bibinfo {volume} {32}} (\bibinfo {year}
  {1971})}\BibitemShut {NoStop}%
\bibitem [{\citenamefont {Schneider}\ and\ \citenamefont
  {Enz}(1971)}]{Schneider:1971}%
  \BibitemOpen
  \bibfield  {author} {\bibinfo {author} {\bibfnamefont {T.}~\bibnamefont
  {Schneider}}\ and\ \bibinfo {author} {\bibfnamefont {C.~P.}\ \bibnamefont
  {Enz}},\ }\href {\doibase 10.1103/PhysRevLett.27.1186} {\bibfield  {journal}
  {\bibinfo  {journal} {Phys. Rev. Lett.}\ }\textbf {\bibinfo {volume} {27}},\
  \bibinfo {pages} {1186} (\bibinfo {year} {1971})}\BibitemShut {NoStop}%
\bibitem [{\citenamefont {Balibar}(2010)}]{Balibar:2010}%
  \BibitemOpen
  \bibfield  {author} {\bibinfo {author} {\bibfnamefont {S.}~\bibnamefont
  {Balibar}},\ }\href {https://www.nature.com/articles/nature08913} {\bibfield
  {journal} {\bibinfo  {journal} {Nature (London)}\ }\textbf {\bibinfo {volume}
  {464}},\ \bibinfo {pages} {176} (\bibinfo {year} {2010})}\BibitemShut
  {NoStop}%
\bibitem [{\citenamefont {Boninsegni}\ and\ \citenamefont
  {Prokof'ev}(2012)}]{Boninsegni:2012}%
  \BibitemOpen
  \bibfield  {author} {\bibinfo {author} {\bibfnamefont {M.}~\bibnamefont
  {Boninsegni}}\ and\ \bibinfo {author} {\bibfnamefont {N.~V.}\ \bibnamefont
  {Prokof'ev}},\ }\href {\doibase 10.1103/RevModPhys.84.759} {\bibfield
  {journal} {\bibinfo  {journal} {Rev. Mod. Phys.}\ }\textbf {\bibinfo {volume}
  {84}},\ \bibinfo {pages} {759} (\bibinfo {year} {2012})}\BibitemShut
  {NoStop}%
\bibitem [{\citenamefont {Lu}\ \emph {et~al.}(2015{\natexlab{a}})\citenamefont
  {Lu}, \citenamefont {Li}, \citenamefont {Petrov},\ and\ \citenamefont
  {Shlyapnikov}}]{Lu2015sds}%
  \BibitemOpen
  \bibfield  {author} {\bibinfo {author} {\bibfnamefont {Z.-K.}\ \bibnamefont
  {Lu}}, \bibinfo {author} {\bibfnamefont {Y.}~\bibnamefont {Li}}, \bibinfo
  {author} {\bibfnamefont {D.~S.}\ \bibnamefont {Petrov}}, \ and\ \bibinfo
  {author} {\bibfnamefont {G.~V.}\ \bibnamefont {Shlyapnikov}},\ }\href
  {\doibase 10.1103/PhysRevLett.115.075303} {\bibfield  {journal} {\bibinfo
  {journal} {Phys. Rev. Lett.}\ }\textbf {\bibinfo {volume} {115}},\ \bibinfo
  {pages} {075303} (\bibinfo {year} {2015}{\natexlab{a}})}\BibitemShut
  {NoStop}%
\bibitem [{\citenamefont {Henkel}\ \emph {et~al.}(2010)\citenamefont {Henkel},
  \citenamefont {Nath},\ and\ \citenamefont {Pohl}}]{Henkel2010tdr}%
  \BibitemOpen
  \bibfield  {author} {\bibinfo {author} {\bibfnamefont {N.}~\bibnamefont
  {Henkel}}, \bibinfo {author} {\bibfnamefont {R.}~\bibnamefont {Nath}}, \ and\
  \bibinfo {author} {\bibfnamefont {T.}~\bibnamefont {Pohl}},\ }\href {\doibase
  10.1103/PhysRevLett.104.195302} {\bibfield  {journal} {\bibinfo  {journal}
  {Phys. Rev. Lett.}\ }\textbf {\bibinfo {volume} {104}},\ \bibinfo {pages}
  {195302} (\bibinfo {year} {2010})}\BibitemShut {NoStop}%
\bibitem [{\citenamefont {Cinti}\ \emph {et~al.}(2010)\citenamefont {Cinti},
  \citenamefont {Jain}, \citenamefont {Boninsegni}, \citenamefont {Micheli},
  \citenamefont {Zoller},\ and\ \citenamefont {Pupillo}}]{Cinti2010sdc}%
  \BibitemOpen
  \bibfield  {author} {\bibinfo {author} {\bibfnamefont {F.}~\bibnamefont
  {Cinti}}, \bibinfo {author} {\bibfnamefont {P.}~\bibnamefont {Jain}},
  \bibinfo {author} {\bibfnamefont {M.}~\bibnamefont {Boninsegni}}, \bibinfo
  {author} {\bibfnamefont {A.}~\bibnamefont {Micheli}}, \bibinfo {author}
  {\bibfnamefont {P.}~\bibnamefont {Zoller}}, \ and\ \bibinfo {author}
  {\bibfnamefont {G.}~\bibnamefont {Pupillo}},\ }\href {\doibase
  10.1103/PhysRevLett.105.135301} {\bibfield  {journal} {\bibinfo  {journal}
  {Phys. Rev. Lett.}\ }\textbf {\bibinfo {volume} {105}},\ \bibinfo {pages}
  {135301} (\bibinfo {year} {2010})}\BibitemShut {NoStop}%
\bibitem [{\citenamefont {Boninsegni}(2012)}]{Boninsegni2012}%
  \BibitemOpen
  \bibfield  {author} {\bibinfo {author} {\bibfnamefont {M.}~\bibnamefont
  {Boninsegni}},\ }\href {\doibase 10.1007/s10909-012-0571-1} {\bibfield
  {journal} {\bibinfo  {journal} {J. Low Temp. Phys.}\ }\textbf {\bibinfo
  {volume} {168}},\ \bibinfo {pages} {137} (\bibinfo {year}
  {2012})}\BibitemShut {NoStop}%
\bibitem [{\citenamefont {L{\'e}onard}\ \emph {et~al.}(2017)\citenamefont
  {L{\'e}onard}, \citenamefont {Morales}, \citenamefont {Zupancic},
  \citenamefont {Esslinger},\ and\ \citenamefont {Donner}}]{Leonard:2017}%
  \BibitemOpen
  \bibfield  {author} {\bibinfo {author} {\bibfnamefont {J.}~\bibnamefont
  {L{\'e}onard}}, \bibinfo {author} {\bibfnamefont {A.}~\bibnamefont
  {Morales}}, \bibinfo {author} {\bibfnamefont {P.}~\bibnamefont {Zupancic}},
  \bibinfo {author} {\bibfnamefont {T.}~\bibnamefont {Esslinger}}, \ and\
  \bibinfo {author} {\bibfnamefont {T.}~\bibnamefont {Donner}},\ }\href
  {https://www.nature.com/articles/nature21067} {\bibfield  {journal} {\bibinfo
   {journal} {Nature (London)}\ }\textbf {\bibinfo {volume} {543}},\ \bibinfo
  {pages} {87} (\bibinfo {year} {2017})}\BibitemShut {NoStop}%
\bibitem [{\citenamefont {Li}\ \emph {et~al.}(2017)\citenamefont {Li},
  \citenamefont {Lee}, \citenamefont {Huang}, \citenamefont {Burchesky},
  \citenamefont {Shteynas}, \citenamefont {Top}, \citenamefont {Jamison},\ and\
  \citenamefont {Ketterle}}]{Li:2017}%
  \BibitemOpen
  \bibfield  {author} {\bibinfo {author} {\bibfnamefont {J.-R.}\ \bibnamefont
  {Li}}, \bibinfo {author} {\bibfnamefont {J.}~\bibnamefont {Lee}}, \bibinfo
  {author} {\bibfnamefont {W.}~\bibnamefont {Huang}}, \bibinfo {author}
  {\bibfnamefont {S.}~\bibnamefont {Burchesky}}, \bibinfo {author}
  {\bibfnamefont {B.}~\bibnamefont {Shteynas}}, \bibinfo {author}
  {\bibfnamefont {F.~{\c{C}}.}\ \bibnamefont {Top}}, \bibinfo {author}
  {\bibfnamefont {A.~O.}\ \bibnamefont {Jamison}}, \ and\ \bibinfo {author}
  {\bibfnamefont {W.}~\bibnamefont {Ketterle}},\ }\href
  {https://www.nature.com/articles/nature21431} {\bibfield  {journal} {\bibinfo
   {journal} {Nature (London)}\ }\textbf {\bibinfo {volume} {543}},\ \bibinfo
  {pages} {91} (\bibinfo {year} {2017})}\BibitemShut {NoStop}%
\bibitem [{\citenamefont {Landau}(1941)}]{Landau41}%
  \BibitemOpen
  \bibfield  {author} {\bibinfo {author} {\bibfnamefont {L.~D.}\ \bibnamefont
  {Landau}},\ }\href@noop {} {\bibfield  {journal} {\bibinfo  {journal} {J.
  Phys. (Moscow)}\ }\textbf {\bibinfo {volume} {5}},\ \bibinfo {pages} {71}
  (\bibinfo {year} {1941})}\BibitemShut {NoStop}%
\bibitem [{\citenamefont {Nozi{\`e}res}(2004)}]{Nozieres2004itr}%
  \BibitemOpen
  \bibfield  {author} {\bibinfo {author} {\bibfnamefont {P.}~\bibnamefont
  {Nozi{\`e}res}},\ }\href {\doibase 10.1023/B:JOLT.0000044234.82957.2f}
  {\bibfield  {journal} {\bibinfo  {journal} {J. Low Temp. Phys.}\ }\textbf
  {\bibinfo {volume} {137}},\ \bibinfo {pages} {45} (\bibinfo {year}
  {2004})}\BibitemShut {NoStop}%
\bibitem [{\citenamefont {O'Dell}\ \emph {et~al.}(2003)\citenamefont {O'Dell},
  \citenamefont {Giovanazzi},\ and\ \citenamefont {Kurizki}}]{ODell2003}%
  \BibitemOpen
  \bibfield  {author} {\bibinfo {author} {\bibfnamefont {D.~H.~J.}\
  \bibnamefont {O'Dell}}, \bibinfo {author} {\bibfnamefont {S.}~\bibnamefont
  {Giovanazzi}}, \ and\ \bibinfo {author} {\bibfnamefont {G.}~\bibnamefont
  {Kurizki}},\ }\href {\doibase 10.1103/PhysRevLett.90.110402} {\bibfield
  {journal} {\bibinfo  {journal} {Phys. Rev. Lett.}\ }\textbf {\bibinfo
  {volume} {90}},\ \bibinfo {pages} {110402} (\bibinfo {year}
  {2003})}\BibitemShut {NoStop}%
\bibitem [{\citenamefont {Santos}\ \emph {et~al.}(2003)\citenamefont {Santos},
  \citenamefont {Shlyapnikov},\ and\ \citenamefont {Lewenstein}}]{Santos:2003}%
  \BibitemOpen
  \bibfield  {author} {\bibinfo {author} {\bibfnamefont {L.}~\bibnamefont
  {Santos}}, \bibinfo {author} {\bibfnamefont {G.~V.}\ \bibnamefont
  {Shlyapnikov}}, \ and\ \bibinfo {author} {\bibfnamefont {M.}~\bibnamefont
  {Lewenstein}},\ }\href {\doibase 10.1103/PhysRevLett.90.250403} {\bibfield
  {journal} {\bibinfo  {journal} {Phys. Rev. Lett.}\ }\textbf {\bibinfo
  {volume} {90}},\ \bibinfo {pages} {250403} (\bibinfo {year}
  {2003})}\BibitemShut {NoStop}%
\bibitem [{\citenamefont {Chomaz}\ \emph {et~al.}(2018)\citenamefont {Chomaz},
  \citenamefont {van Bijnen}, \citenamefont {Petter}, \citenamefont {Faraoni},
  \citenamefont {Baier}, \citenamefont {Becher}, \citenamefont {Mark},
  \citenamefont {W\"achtler}, \citenamefont {Santos},\ and\ \citenamefont
  {Ferlaino}}]{Chomaz:2018}%
  \BibitemOpen
  \bibfield  {author} {\bibinfo {author} {\bibfnamefont {L.}~\bibnamefont
  {Chomaz}}, \bibinfo {author} {\bibfnamefont {R.~M.~W.}\ \bibnamefont {van
  Bijnen}}, \bibinfo {author} {\bibfnamefont {D.}~\bibnamefont {Petter}},
  \bibinfo {author} {\bibfnamefont {G.}~\bibnamefont {Faraoni}}, \bibinfo
  {author} {\bibfnamefont {S.}~\bibnamefont {Baier}}, \bibinfo {author}
  {\bibfnamefont {J.~H.}\ \bibnamefont {Becher}}, \bibinfo {author}
  {\bibfnamefont {M.~J.}\ \bibnamefont {Mark}}, \bibinfo {author}
  {\bibfnamefont {F.}~\bibnamefont {W\"achtler}}, \bibinfo {author}
  {\bibfnamefont {L.}~\bibnamefont {Santos}}, \ and\ \bibinfo {author}
  {\bibfnamefont {F.}~\bibnamefont {Ferlaino}},\ }\href
  {https://www.nature.com/articles/s41567-018-0054-7} {\bibfield  {journal}
  {\bibinfo  {journal} {Nat. Phys.}\ }\textbf {\bibinfo {volume} {14}},\
  \bibinfo {pages} {442} (\bibinfo {year} {2018})}\BibitemShut {NoStop}%
\bibitem [{\citenamefont {Petter}\ \emph {et~al.}(2018)\citenamefont {Petter},
  \citenamefont {Natale}, \citenamefont {van Bijnen}, \citenamefont
  {Patscheider}, \citenamefont {Mark}, \citenamefont {Chomaz},\ and\
  \citenamefont {Ferlaino}}]{Petter:2018}%
  \BibitemOpen
  \bibfield  {author} {\bibinfo {author} {\bibfnamefont {D.}~\bibnamefont
  {Petter}}, \bibinfo {author} {\bibfnamefont {G.}~\bibnamefont {Natale}},
  \bibinfo {author} {\bibfnamefont {R.~M.~W.}\ \bibnamefont {van Bijnen}},
  \bibinfo {author} {\bibfnamefont {A.}~\bibnamefont {Patscheider}}, \bibinfo
  {author} {\bibfnamefont {M.~J.}\ \bibnamefont {Mark}}, \bibinfo {author}
  {\bibfnamefont {L.}~\bibnamefont {Chomaz}}, \ and\ \bibinfo {author}
  {\bibfnamefont {F.}~\bibnamefont {Ferlaino}},\ }\href
  {https://arxiv.org/abs/1811.12115} {\bibfield  {journal} {\bibinfo  {journal}
  {arXiv:1811.12115}\ } (\bibinfo {year} {2018})}\BibitemShut {NoStop}%
\bibitem [{\citenamefont {Kadau}\ \emph {et~al.}(2016)\citenamefont {Kadau},
  \citenamefont {Schmitt}, \citenamefont {Wenzel}, \citenamefont {Wink},
  \citenamefont {Maier}, \citenamefont {Ferrier-Barbut},\ and\ \citenamefont
  {Pfau}}]{Kadau:2016}%
  \BibitemOpen
  \bibfield  {author} {\bibinfo {author} {\bibfnamefont {H.}~\bibnamefont
  {Kadau}}, \bibinfo {author} {\bibfnamefont {M.}~\bibnamefont {Schmitt}},
  \bibinfo {author} {\bibfnamefont {M.}~\bibnamefont {Wenzel}}, \bibinfo
  {author} {\bibfnamefont {C.}~\bibnamefont {Wink}}, \bibinfo {author}
  {\bibfnamefont {T.}~\bibnamefont {Maier}}, \bibinfo {author} {\bibfnamefont
  {I.}~\bibnamefont {Ferrier-Barbut}}, \ and\ \bibinfo {author} {\bibfnamefont
  {T.}~\bibnamefont {Pfau}},\ }\href {http://dx.doi.org/10.1038/nature16485}
  {\bibfield  {journal} {\bibinfo  {journal} {Nature (London)}\ }\textbf
  {\bibinfo {volume} {530}},\ \bibinfo {pages} {194} (\bibinfo {year}
  {2016})}\BibitemShut {NoStop}%
\bibitem [{\citenamefont {Ferrier-Barbut}\ \emph {et~al.}(2016)\citenamefont
  {Ferrier-Barbut}, \citenamefont {Kadau}, \citenamefont {Schmitt},
  \citenamefont {Wenzel},\ and\ \citenamefont {Pfau}}]{Igor:2016}%
  \BibitemOpen
  \bibfield  {author} {\bibinfo {author} {\bibfnamefont {I.}~\bibnamefont
  {Ferrier-Barbut}}, \bibinfo {author} {\bibfnamefont {H.}~\bibnamefont
  {Kadau}}, \bibinfo {author} {\bibfnamefont {M.}~\bibnamefont {Schmitt}},
  \bibinfo {author} {\bibfnamefont {M.}~\bibnamefont {Wenzel}}, \ and\ \bibinfo
  {author} {\bibfnamefont {T.}~\bibnamefont {Pfau}},\ }\href {\doibase
  10.1103/PhysRevLett.116.215301} {\bibfield  {journal} {\bibinfo  {journal}
  {Phys. Rev. Lett.}\ }\textbf {\bibinfo {volume} {116}},\ \bibinfo {pages}
  {215301} (\bibinfo {year} {2016})}\BibitemShut {NoStop}%
\bibitem [{\citenamefont {Chomaz}\ \emph {et~al.}(2016)\citenamefont {Chomaz},
  \citenamefont {Baier}, \citenamefont {Petter}, \citenamefont {Mark},
  \citenamefont {W\"achtler}, \citenamefont {Santos},\ and\ \citenamefont
  {Ferlaino}}]{Chomaz:2016}%
  \BibitemOpen
  \bibfield  {author} {\bibinfo {author} {\bibfnamefont {L.}~\bibnamefont
  {Chomaz}}, \bibinfo {author} {\bibfnamefont {S.}~\bibnamefont {Baier}},
  \bibinfo {author} {\bibfnamefont {D.}~\bibnamefont {Petter}}, \bibinfo
  {author} {\bibfnamefont {M.~J.}\ \bibnamefont {Mark}}, \bibinfo {author}
  {\bibfnamefont {F.}~\bibnamefont {W\"achtler}}, \bibinfo {author}
  {\bibfnamefont {L.}~\bibnamefont {Santos}}, \ and\ \bibinfo {author}
  {\bibfnamefont {F.}~\bibnamefont {Ferlaino}},\ }\href {\doibase
  10.1103/PhysRevX.6.041039} {\bibfield  {journal} {\bibinfo  {journal} {Phys.
  Rev. X}\ }\textbf {\bibinfo {volume} {6}},\ \bibinfo {pages} {041039}
  (\bibinfo {year} {2016})}\BibitemShut {NoStop}%
\bibitem [{\citenamefont {W{\"a}chtler}\ and\ \citenamefont
  {Santos}(2016{\natexlab{a}})}]{Waechtler:2016}%
  \BibitemOpen
  \bibfield  {author} {\bibinfo {author} {\bibfnamefont {F.}~\bibnamefont
  {W{\"a}chtler}}\ and\ \bibinfo {author} {\bibfnamefont {L.}~\bibnamefont
  {Santos}},\ }\href {\doibase 10.1103/PhysRevA.93.061603} {\bibfield
  {journal} {\bibinfo  {journal} {Phys. Rev. A}\ }\textbf {\bibinfo {volume}
  {93}},\ \bibinfo {pages} {061603} (\bibinfo {year}
  {2016}{\natexlab{a}})}\BibitemShut {NoStop}%
\bibitem [{\citenamefont {W{\"a}chtler}\ and\ \citenamefont
  {Santos}(2016{\natexlab{b}})}]{Waechtler:2016b}%
  \BibitemOpen
  \bibfield  {author} {\bibinfo {author} {\bibfnamefont {F.}~\bibnamefont
  {W{\"a}chtler}}\ and\ \bibinfo {author} {\bibfnamefont {L.}~\bibnamefont
  {Santos}},\ }\href {\doibase 10.1103/PhysRevA.94.043618} {\bibfield
  {journal} {\bibinfo  {journal} {Phys. Rev. A}\ }\textbf {\bibinfo {volume}
  {94}},\ \bibinfo {pages} {043618} (\bibinfo {year}
  {2016}{\natexlab{b}})}\BibitemShut {NoStop}%
\bibitem [{\citenamefont {Schmitt}\ \emph {et~al.}(2016)\citenamefont
  {Schmitt}, \citenamefont {Wenzel}, \citenamefont {B{\"o}ttcher},
  \citenamefont {Ferrier-Barbut},\ and\ \citenamefont {Pfau}}]{Schmitt:2016}%
  \BibitemOpen
  \bibfield  {author} {\bibinfo {author} {\bibfnamefont {M.}~\bibnamefont
  {Schmitt}}, \bibinfo {author} {\bibfnamefont {M.}~\bibnamefont {Wenzel}},
  \bibinfo {author} {\bibfnamefont {F.}~\bibnamefont {B{\"o}ttcher}}, \bibinfo
  {author} {\bibfnamefont {I.}~\bibnamefont {Ferrier-Barbut}}, \ and\ \bibinfo
  {author} {\bibfnamefont {T.}~\bibnamefont {Pfau}},\ }\href
  {https://www.nature.com/articles/nature20126} {\bibfield  {journal} {\bibinfo
   {journal} {Nature (London)}\ }\textbf {\bibinfo {volume} {539}},\ \bibinfo
  {pages} {259} (\bibinfo {year} {2016})}\BibitemShut {NoStop}%
\bibitem [{\citenamefont {Ferrier-Barbut}\ \emph
  {et~al.}(2018{\natexlab{a}})\citenamefont {Ferrier-Barbut}, \citenamefont
  {Wenzel}, \citenamefont {Schmitt}, \citenamefont {B\"ottcher},\ and\
  \citenamefont {Pfau}}]{Igor2018ooa}%
  \BibitemOpen
  \bibfield  {author} {\bibinfo {author} {\bibfnamefont {I.}~\bibnamefont
  {Ferrier-Barbut}}, \bibinfo {author} {\bibfnamefont {M.}~\bibnamefont
  {Wenzel}}, \bibinfo {author} {\bibfnamefont {M.}~\bibnamefont {Schmitt}},
  \bibinfo {author} {\bibfnamefont {F.}~\bibnamefont {B\"ottcher}}, \ and\
  \bibinfo {author} {\bibfnamefont {T.}~\bibnamefont {Pfau}},\ }\href {\doibase
  10.1103/PhysRevA.97.011604} {\bibfield  {journal} {\bibinfo  {journal} {Phys.
  Rev. A}\ }\textbf {\bibinfo {volume} {97}},\ \bibinfo {pages} {011604}
  (\bibinfo {year} {2018}{\natexlab{a}})}\BibitemShut {NoStop}%
\bibitem [{\citenamefont {Bisset}\ \emph {et~al.}(2016)\citenamefont {Bisset},
  \citenamefont {Wilson}, \citenamefont {Baillie},\ and\ \citenamefont
  {Blakie}}]{Bisset:2016}%
  \BibitemOpen
  \bibfield  {author} {\bibinfo {author} {\bibfnamefont {R.~N.}\ \bibnamefont
  {Bisset}}, \bibinfo {author} {\bibfnamefont {R.~M.}\ \bibnamefont {Wilson}},
  \bibinfo {author} {\bibfnamefont {D.}~\bibnamefont {Baillie}}, \ and\
  \bibinfo {author} {\bibfnamefont {P.~B.}\ \bibnamefont {Blakie}},\ }\href
  {\doibase 10.1103/PhysRevA.94.033619} {\bibfield  {journal} {\bibinfo
  {journal} {Phys. Rev. A}\ }\textbf {\bibinfo {volume} {94}},\ \bibinfo
  {pages} {033619} (\bibinfo {year} {2016})}\BibitemShut {NoStop}%
\bibitem [{\citenamefont {Wenzel}\ \emph {et~al.}(2017)\citenamefont {Wenzel},
  \citenamefont {B\"ottcher}, \citenamefont {Langen}, \citenamefont
  {Ferrier-Barbut},\ and\ \citenamefont {Pfau}}]{Wenzel:2017}%
  \BibitemOpen
  \bibfield  {author} {\bibinfo {author} {\bibfnamefont {M.}~\bibnamefont
  {Wenzel}}, \bibinfo {author} {\bibfnamefont {F.}~\bibnamefont {B\"ottcher}},
  \bibinfo {author} {\bibfnamefont {T.}~\bibnamefont {Langen}}, \bibinfo
  {author} {\bibfnamefont {I.}~\bibnamefont {Ferrier-Barbut}}, \ and\ \bibinfo
  {author} {\bibfnamefont {T.}~\bibnamefont {Pfau}},\ }\href {\doibase
  10.1103/PhysRevA.96.053630} {\bibfield  {journal} {\bibinfo  {journal} {Phys.
  Rev. A}\ }\textbf {\bibinfo {volume} {96}},\ \bibinfo {pages} {053630}
  (\bibinfo {year} {2017})}\BibitemShut {NoStop}%
\bibitem [{\citenamefont {Baillie}\ and\ \citenamefont
  {Blakie}(2018)}]{Baillie:2018}%
  \BibitemOpen
  \bibfield  {author} {\bibinfo {author} {\bibfnamefont {D.}~\bibnamefont
  {Baillie}}\ and\ \bibinfo {author} {\bibfnamefont {P.~B.}\ \bibnamefont
  {Blakie}},\ }\href {\doibase 10.1103/PhysRevLett.121.195301} {\bibfield
  {journal} {\bibinfo  {journal} {Phys. Rev. Lett.}\ }\textbf {\bibinfo
  {volume} {121}},\ \bibinfo {pages} {195301} (\bibinfo {year}
  {2018})}\BibitemShut {NoStop}%
\bibitem [{\citenamefont {Petrov}(2015{\natexlab{a}})}]{petrov2015}%
  \BibitemOpen
  \bibfield  {author} {\bibinfo {author} {\bibfnamefont {D.}~\bibnamefont
  {Petrov}},\ }\href {https://link.aps.org/doi/10.1103/PhysRevLett.115.155302}
  {\bibfield  {journal} {\bibinfo  {journal} {Phys. Rev. Lett.}\ }\textbf
  {\bibinfo {volume} {115}},\ \bibinfo {pages} {155302} (\bibinfo {year}
  {2015}{\natexlab{a}})}\BibitemShut {NoStop}%
\bibitem [{\citenamefont {Cabrera}\ \emph {et~al.}(2018)\citenamefont
  {Cabrera}, \citenamefont {Tanzi}, \citenamefont {Sanz}, \citenamefont
  {Naylor}, \citenamefont {Thomas}, \citenamefont {Cheiney},\ and\
  \citenamefont {Tarruell}}]{Cabrera2018}%
  \BibitemOpen
  \bibfield  {author} {\bibinfo {author} {\bibfnamefont {C.~R.}\ \bibnamefont
  {Cabrera}}, \bibinfo {author} {\bibfnamefont {L.}~\bibnamefont {Tanzi}},
  \bibinfo {author} {\bibfnamefont {J.}~\bibnamefont {Sanz}}, \bibinfo {author}
  {\bibfnamefont {B.}~\bibnamefont {Naylor}}, \bibinfo {author} {\bibfnamefont
  {P.}~\bibnamefont {Thomas}}, \bibinfo {author} {\bibfnamefont
  {P.}~\bibnamefont {Cheiney}}, \ and\ \bibinfo {author} {\bibfnamefont
  {L.}~\bibnamefont {Tarruell}},\ }\href {\doibase 10.1126/science.aao5686}
  {\bibfield  {journal} {\bibinfo  {journal} {Science}\ }\textbf {\bibinfo
  {volume} {359}},\ \bibinfo {pages} {301} (\bibinfo {year}
  {2018})}\BibitemShut {NoStop}%
\bibitem [{\citenamefont {Semeghini}\ \emph {et~al.}(2018)\citenamefont
  {Semeghini}, \citenamefont {Ferioli}, \citenamefont {Masi}, \citenamefont
  {Mazzinghi}, \citenamefont {Wolswijk}, \citenamefont {Minardi}, \citenamefont
  {Modugno}, \citenamefont {Modugno}, \citenamefont {Inguscio},\ and\
  \citenamefont {Fattori}}]{semeghini2018}%
  \BibitemOpen
  \bibfield  {author} {\bibinfo {author} {\bibfnamefont {G.}~\bibnamefont
  {Semeghini}}, \bibinfo {author} {\bibfnamefont {G.}~\bibnamefont {Ferioli}},
  \bibinfo {author} {\bibfnamefont {L.}~\bibnamefont {Masi}}, \bibinfo {author}
  {\bibfnamefont {C.}~\bibnamefont {Mazzinghi}}, \bibinfo {author}
  {\bibfnamefont {L.}~\bibnamefont {Wolswijk}}, \bibinfo {author}
  {\bibfnamefont {F.}~\bibnamefont {Minardi}}, \bibinfo {author} {\bibfnamefont
  {M.}~\bibnamefont {Modugno}}, \bibinfo {author} {\bibfnamefont
  {G.}~\bibnamefont {Modugno}}, \bibinfo {author} {\bibfnamefont
  {M.}~\bibnamefont {Inguscio}}, \ and\ \bibinfo {author} {\bibfnamefont
  {M.}~\bibnamefont {Fattori}},\ }\href
  {https://link.aps.org/doi/10.1103/PhysRevLett.120.235301} {\bibfield
  {journal} {\bibinfo  {journal} {Phys. Rev. Lett.}\ }\textbf {\bibinfo
  {volume} {120}},\ \bibinfo {pages} {235301} (\bibinfo {year}
  {2018})}\BibitemShut {NoStop}%
\bibitem [{\citenamefont {Cheiney}\ \emph {et~al.}(2018)\citenamefont
  {Cheiney}, \citenamefont {Cabrera}, \citenamefont {Sanz}, \citenamefont
  {Naylor}, \citenamefont {Tanzi},\ and\ \citenamefont
  {Tarruell}}]{cheiney2018}%
  \BibitemOpen
  \bibfield  {author} {\bibinfo {author} {\bibfnamefont {P.}~\bibnamefont
  {Cheiney}}, \bibinfo {author} {\bibfnamefont {C.}~\bibnamefont {Cabrera}},
  \bibinfo {author} {\bibfnamefont {J.}~\bibnamefont {Sanz}}, \bibinfo {author}
  {\bibfnamefont {B.}~\bibnamefont {Naylor}}, \bibinfo {author} {\bibfnamefont
  {L.}~\bibnamefont {Tanzi}}, \ and\ \bibinfo {author} {\bibfnamefont
  {L.}~\bibnamefont {Tarruell}},\ }\href
  {https://link.aps.org/doi/10.1103/PhysRevLett.120.135301} {\bibfield
  {journal} {\bibinfo  {journal} {Phys. Rev. Lett.}\ }\textbf {\bibinfo
  {volume} {120}},\ \bibinfo {pages} {135301} (\bibinfo {year}
  {2018})}\BibitemShut {NoStop}%
\bibitem [{\citenamefont {Roccuzzo}\ and\ \citenamefont
  {Ancilotto}(2018)}]{Ancilotto:2018}%
  \BibitemOpen
  \bibfield  {author} {\bibinfo {author} {\bibfnamefont {S.~M.}\ \bibnamefont
  {Roccuzzo}}\ and\ \bibinfo {author} {\bibfnamefont {F.}~\bibnamefont
  {Ancilotto}},\ }\href {https://arxiv.org/abs/1810.12229} {\bibfield
  {journal} {\bibinfo  {journal} {arXiv:1810.12229}\ } (\bibinfo {year}
  {2018})}\BibitemShut {NoStop}%
\bibitem [{\citenamefont {Tanzi}\ \emph {et~al.}(2018)\citenamefont {Tanzi},
  \citenamefont {Lucioni}, \citenamefont {Fama}, \citenamefont {Catani},
  \citenamefont {Fioretti}, \citenamefont {Gabbanini}, \citenamefont {Bisset},
  \citenamefont {Santos},\ and\ \citenamefont {Modugno}}]{Tanzi:2018}%
  \BibitemOpen
  \bibfield  {author} {\bibinfo {author} {\bibfnamefont {L.}~\bibnamefont
  {Tanzi}}, \bibinfo {author} {\bibfnamefont {E.}~\bibnamefont {Lucioni}},
  \bibinfo {author} {\bibfnamefont {F.}~\bibnamefont {Fama}}, \bibinfo {author}
  {\bibfnamefont {J.}~\bibnamefont {Catani}}, \bibinfo {author} {\bibfnamefont
  {A.}~\bibnamefont {Fioretti}}, \bibinfo {author} {\bibfnamefont
  {C.}~\bibnamefont {Gabbanini}}, \bibinfo {author} {\bibfnamefont {R.~N.}\
  \bibnamefont {Bisset}}, \bibinfo {author} {\bibfnamefont {L.}~\bibnamefont
  {Santos}}, \ and\ \bibinfo {author} {\bibfnamefont {G.}~\bibnamefont
  {Modugno}},\ }\href {https://arxiv.org/abs/1811.02613} {\bibfield  {journal}
  {\bibinfo  {journal} {arXiv:1811.02613}\ } (\bibinfo {year}
  {2018})}\BibitemShut {NoStop}%
\bibitem [{\citenamefont {B{\"o}ttcher}\ \emph {et~al.}(2019)\citenamefont
  {B{\"o}ttcher}, \citenamefont {Schmidt}, \citenamefont {Wenzel},
  \citenamefont {Hertkorn}, \citenamefont {Guo}, \citenamefont {Langen},\ and\
  \citenamefont {Pfau}}]{Bottcher2019}%
  \BibitemOpen
  \bibfield  {author} {\bibinfo {author} {\bibfnamefont {F.}~\bibnamefont
  {B{\"o}ttcher}}, \bibinfo {author} {\bibfnamefont {J.-N.}\ \bibnamefont
  {Schmidt}}, \bibinfo {author} {\bibfnamefont {M.}~\bibnamefont {Wenzel}},
  \bibinfo {author} {\bibfnamefont {J.}~\bibnamefont {Hertkorn}}, \bibinfo
  {author} {\bibfnamefont {M.}~\bibnamefont {Guo}}, \bibinfo {author}
  {\bibfnamefont {T.}~\bibnamefont {Langen}}, \ and\ \bibinfo {author}
  {\bibfnamefont {T.}~\bibnamefont {Pfau}},\ }\href
  {https://arxiv.org/abs/1901.07982} {\bibfield  {journal} {\bibinfo  {journal}
  {arXiv:1901.07982}\ } (\bibinfo {year} {2019})}\BibitemShut {NoStop}%
\bibitem [{\citenamefont {Aikawa}\ \emph {et~al.}(2012)\citenamefont {Aikawa},
  \citenamefont {Frisch}, \citenamefont {Mark}, \citenamefont {Baier},
  \citenamefont {Rietzler}, \citenamefont {Grimm},\ and\ \citenamefont
  {Ferlaino}}]{Aikawa:2012}%
  \BibitemOpen
  \bibfield  {author} {\bibinfo {author} {\bibfnamefont {K.}~\bibnamefont
  {Aikawa}}, \bibinfo {author} {\bibfnamefont {A.}~\bibnamefont {Frisch}},
  \bibinfo {author} {\bibfnamefont {M.}~\bibnamefont {Mark}}, \bibinfo {author}
  {\bibfnamefont {S.}~\bibnamefont {Baier}}, \bibinfo {author} {\bibfnamefont
  {A.}~\bibnamefont {Rietzler}}, \bibinfo {author} {\bibfnamefont
  {R.}~\bibnamefont {Grimm}}, \ and\ \bibinfo {author} {\bibfnamefont
  {F.}~\bibnamefont {Ferlaino}},\ }\href {\doibase
  10.1103/PhysRevLett.108.210401} {\bibfield  {journal} {\bibinfo  {journal}
  {Phys. Rev. Lett.}\ }\textbf {\bibinfo {volume} {108}},\ \bibinfo {pages}
  {210401} (\bibinfo {year} {2012})}\BibitemShut {NoStop}%
\bibitem [{\citenamefont {Trautmann}\ \emph {et~al.}(2018)\citenamefont
  {Trautmann}, \citenamefont {Ilzh\"ofer}, \citenamefont {Durastante},
  \citenamefont {Politi}, \citenamefont {Sohmen}, \citenamefont {Mark},\ and\
  \citenamefont {Ferlaino}}]{Trautmann2018}%
  \BibitemOpen
  \bibfield  {author} {\bibinfo {author} {\bibfnamefont {A.}~\bibnamefont
  {Trautmann}}, \bibinfo {author} {\bibfnamefont {P.}~\bibnamefont
  {Ilzh\"ofer}}, \bibinfo {author} {\bibfnamefont {G.}~\bibnamefont
  {Durastante}}, \bibinfo {author} {\bibfnamefont {C.}~\bibnamefont {Politi}},
  \bibinfo {author} {\bibfnamefont {M.}~\bibnamefont {Sohmen}}, \bibinfo
  {author} {\bibfnamefont {M.~J.}\ \bibnamefont {Mark}}, \ and\ \bibinfo
  {author} {\bibfnamefont {F.}~\bibnamefont {Ferlaino}},\ }\href {\doibase
  10.1103/PhysRevLett.121.213601} {\bibfield  {journal} {\bibinfo  {journal}
  {Phys. Rev. Lett.}\ }\textbf {\bibinfo {volume} {121}},\ \bibinfo {pages}
  {213601} (\bibinfo {year} {2018})}\BibitemShut {NoStop}%
\bibitem [{\citenamefont {Baier}\ \emph {et~al.}(2016)\citenamefont {Baier},
  \citenamefont {Mark}, \citenamefont {Petter}, \citenamefont {Aikawa},
  \citenamefont {Chomaz}, \citenamefont {Cai}, \citenamefont {Baranov},
  \citenamefont {Zoller},\ and\ \citenamefont {Ferlaino}}]{Baier2016ebh}%
  \BibitemOpen
  \bibfield  {author} {\bibinfo {author} {\bibfnamefont {S.}~\bibnamefont
  {Baier}}, \bibinfo {author} {\bibfnamefont {M.~J.}\ \bibnamefont {Mark}},
  \bibinfo {author} {\bibfnamefont {D.}~\bibnamefont {Petter}}, \bibinfo
  {author} {\bibfnamefont {K.}~\bibnamefont {Aikawa}}, \bibinfo {author}
  {\bibfnamefont {L.}~\bibnamefont {Chomaz}}, \bibinfo {author} {\bibfnamefont
  {Z.}~\bibnamefont {Cai}}, \bibinfo {author} {\bibfnamefont {M.}~\bibnamefont
  {Baranov}}, \bibinfo {author} {\bibfnamefont {P.}~\bibnamefont {Zoller}}, \
  and\ \bibinfo {author} {\bibfnamefont {F.}~\bibnamefont {Ferlaino}},\ }\href
  {\doibase 10.1126/science.aac9812} {\bibfield  {journal} {\bibinfo  {journal}
  {Science}\ }\textbf {\bibinfo {volume} {352}},\ \bibinfo {pages} {201}
  (\bibinfo {year} {2016})}\BibitemShut {NoStop}%
\bibitem [{sup()}]{supmat}%
  \BibitemOpen
  \href@noop {} {\emph {\bibinfo {title} {\textnormal{See Supplemental Material
  at [URL] for details on the experimental setup, measurement schemes, analysis
  and theory calculations.}}}}\BibitemShut {Stop}%
\bibitem [{\citenamefont {Ferrier-Barbut}\ \emph
  {et~al.}(2018{\natexlab{b}})\citenamefont {Ferrier-Barbut}, \citenamefont
  {Wenzel}, \citenamefont {B\"ottcher}, \citenamefont {Langen}, \citenamefont
  {Isoard}, \citenamefont {Stringari},\ and\ \citenamefont {Pfau}}]{Igor:2018}%
  \BibitemOpen
  \bibfield  {author} {\bibinfo {author} {\bibfnamefont {I.}~\bibnamefont
  {Ferrier-Barbut}}, \bibinfo {author} {\bibfnamefont {M.}~\bibnamefont
  {Wenzel}}, \bibinfo {author} {\bibfnamefont {F.}~\bibnamefont {B\"ottcher}},
  \bibinfo {author} {\bibfnamefont {T.}~\bibnamefont {Langen}}, \bibinfo
  {author} {\bibfnamefont {M.}~\bibnamefont {Isoard}}, \bibinfo {author}
  {\bibfnamefont {S.}~\bibnamefont {Stringari}}, \ and\ \bibinfo {author}
  {\bibfnamefont {T.}~\bibnamefont {Pfau}},\ }\href {\doibase
  10.1103/PhysRevLett.120.160402} {\bibfield  {journal} {\bibinfo  {journal}
  {Phys. Rev. Lett.}\ }\textbf {\bibinfo {volume} {120}},\ \bibinfo {pages}
  {160402} (\bibinfo {year} {2018}{\natexlab{b}})}\BibitemShut {NoStop}%
\bibitem [{\citenamefont {Gammal}\ \emph {et~al.}(2000)\citenamefont {Gammal},
  \citenamefont {Frederico}, \citenamefont {Tomio},\ and\ \citenamefont
  {Chomaz}}]{Gammal:2000}%
  \BibitemOpen
  \bibfield  {author} {\bibinfo {author} {\bibfnamefont {A.}~\bibnamefont
  {Gammal}}, \bibinfo {author} {\bibfnamefont {T.}~\bibnamefont {Frederico}},
  \bibinfo {author} {\bibfnamefont {L.}~\bibnamefont {Tomio}}, \ and\ \bibinfo
  {author} {\bibfnamefont {P.}~\bibnamefont {Chomaz}},\ }\href {\doibase
  10.1088/0953-4075/33/19/316} {\bibfield  {journal} {\bibinfo  {journal} {J.
  Phys. B}\ }\textbf {\bibinfo {volume} {33}},\ \bibinfo {pages} {4053}
  (\bibinfo {year} {2000})}\BibitemShut {NoStop}%
\bibitem [{\citenamefont {Bulgac}(2002)}]{Bulgac:2002}%
  \BibitemOpen
  \bibfield  {author} {\bibinfo {author} {\bibfnamefont {A.}~\bibnamefont
  {Bulgac}},\ }\href {\doibase 10.1103/PhysRevLett.89.050402} {\bibfield
  {journal} {\bibinfo  {journal} {Phys. Rev. Lett.}\ }\textbf {\bibinfo
  {volume} {89}},\ \bibinfo {pages} {050402} (\bibinfo {year}
  {2002})}\BibitemShut {NoStop}%
\bibitem [{\citenamefont {Petrov}(2015{\natexlab{b}})}]{Petrov:2015}%
  \BibitemOpen
  \bibfield  {author} {\bibinfo {author} {\bibfnamefont {D.~S.}\ \bibnamefont
  {Petrov}},\ }\href {\doibase 10.1103/PhysRevLett.115.155302} {\bibfield
  {journal} {\bibinfo  {journal} {Phys. Rev. Lett.}\ }\textbf {\bibinfo
  {volume} {115}},\ \bibinfo {pages} {155302} (\bibinfo {year}
  {2015}{\natexlab{b}})}\BibitemShut {NoStop}%
\bibitem [{\citenamefont {W\"achtler}\ and\ \citenamefont
  {Santos}(2016)}]{Waechtler1:2016b}%
  \BibitemOpen
  \bibfield  {author} {\bibinfo {author} {\bibfnamefont {F.}~\bibnamefont
  {W\"achtler}}\ and\ \bibinfo {author} {\bibfnamefont {L.}~\bibnamefont
  {Santos}},\ }\href {\doibase 10.1103/PhysRevA.94.043618} {\bibfield
  {journal} {\bibinfo  {journal} {Phys. Rev. A}\ }\textbf {\bibinfo {volume}
  {94}},\ \bibinfo {pages} {043618} (\bibinfo {year} {2016})}\BibitemShut
  {NoStop}%
\bibitem [{\citenamefont {Wenzel}\ \emph {et~al.}(2018)\citenamefont {Wenzel},
  \citenamefont {B\"ottcher}, \citenamefont {Schmidt}, \citenamefont
  {Eisenmann}, \citenamefont {Langen}, \citenamefont {Pfau},\ and\
  \citenamefont {Ferrier-Barbut}}]{Wenzel:2018}%
  \BibitemOpen
  \bibfield  {author} {\bibinfo {author} {\bibfnamefont {M.}~\bibnamefont
  {Wenzel}}, \bibinfo {author} {\bibfnamefont {F.}~\bibnamefont {B\"ottcher}},
  \bibinfo {author} {\bibfnamefont {J.-N.}\ \bibnamefont {Schmidt}}, \bibinfo
  {author} {\bibfnamefont {M.}~\bibnamefont {Eisenmann}}, \bibinfo {author}
  {\bibfnamefont {T.}~\bibnamefont {Langen}}, \bibinfo {author} {\bibfnamefont
  {T.}~\bibnamefont {Pfau}}, \ and\ \bibinfo {author} {\bibfnamefont
  {I.}~\bibnamefont {Ferrier-Barbut}},\ }\href {\doibase
  10.1103/PhysRevLett.121.030401} {\bibfield  {journal} {\bibinfo  {journal}
  {Phys. Rev. Lett.}\ }\textbf {\bibinfo {volume} {121}},\ \bibinfo {pages}
  {030401} (\bibinfo {year} {2018})}\BibitemShut {NoStop}%
\bibitem [{\citenamefont {Josephson}(1962)}]{Josephson:1962}%
  \BibitemOpen
  \bibfield  {author} {\bibinfo {author} {\bibfnamefont {B.}~\bibnamefont
  {Josephson}},\ }\href {\doibase https://doi.org/10.1016/0031-9163(62)91369-0}
  {\bibfield  {journal} {\bibinfo  {journal} {Phys. Lett.}\ }\textbf {\bibinfo
  {volume} {1}},\ \bibinfo {pages} {251 } (\bibinfo {year} {1962})}\BibitemShut
  {NoStop}%
\bibitem [{\citenamefont {Javanainen}(1986)}]{Javanainen:1986}%
  \BibitemOpen
  \bibfield  {author} {\bibinfo {author} {\bibfnamefont {J.}~\bibnamefont
  {Javanainen}},\ }\href {\doibase 10.1103/PhysRevLett.57.3164} {\bibfield
  {journal} {\bibinfo  {journal} {Phys. Rev. Lett.}\ }\textbf {\bibinfo
  {volume} {57}},\ \bibinfo {pages} {3164} (\bibinfo {year}
  {1986})}\BibitemShut {NoStop}%
\bibitem [{\citenamefont {Raghavan}\ \emph {et~al.}(1999)\citenamefont
  {Raghavan}, \citenamefont {Smerzi}, \citenamefont {Fantoni},\ and\
  \citenamefont {Shenoy}}]{Smerzi}%
  \BibitemOpen
  \bibfield  {author} {\bibinfo {author} {\bibfnamefont {S.}~\bibnamefont
  {Raghavan}}, \bibinfo {author} {\bibfnamefont {A.}~\bibnamefont {Smerzi}},
  \bibinfo {author} {\bibfnamefont {S.}~\bibnamefont {Fantoni}}, \ and\
  \bibinfo {author} {\bibfnamefont {S.~R.}\ \bibnamefont {Shenoy}},\ }\href
  {\doibase 10.1103/PhysRevA.59.620} {\bibfield  {journal} {\bibinfo  {journal}
  {Phys. Rev. A}\ }\textbf {\bibinfo {volume} {59}},\ \bibinfo {pages} {620}
  (\bibinfo {year} {1999})}\BibitemShut {NoStop}%
\bibitem [{foo({\natexlab{a}})}]{footnoteSD}%
  \BibitemOpen
  \href@noop {} {\emph {\bibinfo {title} {\textnormal{A similar behavior is
  also seen for $\Dy$ for $N<4000$ (out of the scale of
  Fig.\,\ref{fig:theory}(g)).}}}} ({\natexlab{a}})\BibitemShut {NoStop}%
\bibitem [{foo({\natexlab{b}})}]{footnotewings}%
  \BibitemOpen
  \href@noop {} {\emph {\bibinfo {title} {\textnormal{We note that a similar
  behavior is also found in our $\Er$ phase diagram but for larger $N$ than the
  one reported here.}}}} ({\natexlab{b}})\BibitemShut {NoStop}%
\bibitem [{\citenamefont {Tang}\ \emph {et~al.}(2015)\citenamefont {Tang},
  \citenamefont {Sykes}, \citenamefont {Burdick}, \citenamefont {Bohn},\ and\
  \citenamefont {Lev}}]{Tang2015}%
  \BibitemOpen
  \bibfield  {author} {\bibinfo {author} {\bibfnamefont {Y.}~\bibnamefont
  {Tang}}, \bibinfo {author} {\bibfnamefont {A.}~\bibnamefont {Sykes}},
  \bibinfo {author} {\bibfnamefont {N.~Q.}\ \bibnamefont {Burdick}}, \bibinfo
  {author} {\bibfnamefont {J.~L.}\ \bibnamefont {Bohn}}, \ and\ \bibinfo
  {author} {\bibfnamefont {B.~L.}\ \bibnamefont {Lev}},\ }\href {\doibase
  10.1103/PhysRevA.92.022703} {\bibfield  {journal} {\bibinfo  {journal} {Phys.
  Rev. A}\ }\textbf {\bibinfo {volume} {92}},\ \bibinfo {pages} {022703}
  (\bibinfo {year} {2015})}\BibitemShut {NoStop}%
\bibitem [{\citenamefont {Greiner}\ \emph {et~al.}(2002)\citenamefont
  {Greiner}, \citenamefont {Mandel}, \citenamefont {Esslinger}, \citenamefont
  {H{\"a}nsch},\ and\ \citenamefont {Bloch}}]{Greiner2002}%
  \BibitemOpen
  \bibfield  {author} {\bibinfo {author} {\bibfnamefont {M.}~\bibnamefont
  {Greiner}}, \bibinfo {author} {\bibfnamefont {O.}~\bibnamefont {Mandel}},
  \bibinfo {author} {\bibfnamefont {T.}~\bibnamefont {Esslinger}}, \bibinfo
  {author} {\bibfnamefont {T.~W.}\ \bibnamefont {H{\"a}nsch}}, \ and\ \bibinfo
  {author} {\bibfnamefont {I.}~\bibnamefont {Bloch}},\ }\href
  {https://www.nature.com/articles/415039a} {\bibfield  {journal} {\bibinfo
  {journal} {Nature (London)}\ }\textbf {\bibinfo {volume} {415}},\ \bibinfo
  {pages} {39} (\bibinfo {year} {2002})}\BibitemShut {NoStop}%
\bibitem [{\citenamefont {Greiner}\ \emph {et~al.}(2001)\citenamefont
  {Greiner}, \citenamefont {Bloch}, \citenamefont {Mandel}, \citenamefont
  {H\"ansch},\ and\ \citenamefont {Esslinger}}]{Greiner2001}%
  \BibitemOpen
  \bibfield  {author} {\bibinfo {author} {\bibfnamefont {M.}~\bibnamefont
  {Greiner}}, \bibinfo {author} {\bibfnamefont {I.}~\bibnamefont {Bloch}},
  \bibinfo {author} {\bibfnamefont {O.}~\bibnamefont {Mandel}}, \bibinfo
  {author} {\bibfnamefont {T.~W.}\ \bibnamefont {H\"ansch}}, \ and\ \bibinfo
  {author} {\bibfnamefont {T.}~\bibnamefont {Esslinger}},\ }\href {\doibase
  10.1103/PhysRevLett.87.160405} {\bibfield  {journal} {\bibinfo  {journal}
  {Phys. Rev. Lett.}\ }\textbf {\bibinfo {volume} {87}},\ \bibinfo {pages}
  {160405} (\bibinfo {year} {2001})}\BibitemShut {NoStop}%
\bibitem [{\citenamefont {Paredes}\ \emph {et~al.}(2004)\citenamefont
  {Paredes}, \citenamefont {Widera}, \citenamefont {Murg}, \citenamefont
  {Mandel}, \citenamefont {F{\"o}lling}, \citenamefont {Cirac}, \citenamefont
  {Shlyapnikov}, \citenamefont {H{\"a}nsch},\ and\ \citenamefont
  {Bloch}}]{Paredes2004}%
  \BibitemOpen
  \bibfield  {author} {\bibinfo {author} {\bibfnamefont {B.}~\bibnamefont
  {Paredes}}, \bibinfo {author} {\bibfnamefont {A.}~\bibnamefont {Widera}},
  \bibinfo {author} {\bibfnamefont {V.}~\bibnamefont {Murg}}, \bibinfo {author}
  {\bibfnamefont {O.}~\bibnamefont {Mandel}}, \bibinfo {author} {\bibfnamefont
  {S.}~\bibnamefont {F{\"o}lling}}, \bibinfo {author} {\bibfnamefont
  {I.}~\bibnamefont {Cirac}}, \bibinfo {author} {\bibfnamefont {G.~V.}\
  \bibnamefont {Shlyapnikov}}, \bibinfo {author} {\bibfnamefont {T.~W.}\
  \bibnamefont {H{\"a}nsch}}, \ and\ \bibinfo {author} {\bibfnamefont
  {I.}~\bibnamefont {Bloch}},\ }\href
  {https://www.nature.com/articles/nature02530} {\bibfield  {journal} {\bibinfo
   {journal} {Nature (London)}\ }\textbf {\bibinfo {volume} {429}},\ \bibinfo
  {pages} {277} (\bibinfo {year} {2004})}\BibitemShut {NoStop}%
\bibitem [{\citenamefont {Hadzibabic}\ \emph {et~al.}(2004)\citenamefont
  {Hadzibabic}, \citenamefont {Stock}, \citenamefont {Battelier}, \citenamefont
  {Bretin},\ and\ \citenamefont {Dalibard}}]{Hadzibabic}%
  \BibitemOpen
  \bibfield  {author} {\bibinfo {author} {\bibfnamefont {Z.}~\bibnamefont
  {Hadzibabic}}, \bibinfo {author} {\bibfnamefont {S.}~\bibnamefont {Stock}},
  \bibinfo {author} {\bibfnamefont {B.}~\bibnamefont {Battelier}}, \bibinfo
  {author} {\bibfnamefont {V.}~\bibnamefont {Bretin}}, \ and\ \bibinfo {author}
  {\bibfnamefont {J.}~\bibnamefont {Dalibard}},\ }\href {\doibase
  10.1103/PhysRevLett.93.180403} {\bibfield  {journal} {\bibinfo  {journal}
  {Phys. Rev. Lett.}\ }\textbf {\bibinfo {volume} {93}},\ \bibinfo {pages}
  {180403} (\bibinfo {year} {2004})}\BibitemShut {NoStop}%
\bibitem [{Not({\natexlab{a}})}]{Noteaberation}%
  \BibitemOpen
  \href@noop {} {\emph {\bibinfo {title} {\textnormal{In our absorption imaging
  a residual asymmetry can be observed between $\pm k_y$. This asymmetry,
  providing a sharper peak at $k_y>0$ stems from optical aberrations in our
  imperfect optical setup. These imperfections make the quantitative analysis
  of our data more challenging, yet they do not affect the quantitative
  interpretation. In particular we have checked that the peak structures are
  also observable along a distinct imaging axis, non-orthogonal to $y$, using
  an independent imaging setup.}}}} ({\natexlab{a}})\BibitemShut {NoStop}%
\bibitem [{\citenamefont {Takeda}\ \emph {et~al.}(1982)\citenamefont {Takeda},
  \citenamefont {Ina},\ and\ \citenamefont {Kobayashi}}]{takeda1982ftm}%
  \BibitemOpen
  \bibfield  {author} {\bibinfo {author} {\bibfnamefont {M.}~\bibnamefont
  {Takeda}}, \bibinfo {author} {\bibfnamefont {H.}~\bibnamefont {Ina}}, \ and\
  \bibinfo {author} {\bibfnamefont {S.}~\bibnamefont {Kobayashi}},\ }\href
  {http://www.osapublishing.org/abstract.cfm?URI=josa-72-1-156} {\bibfield
  {journal} {\bibinfo  {journal} {J. Opt. Soc. Am.}\ }\textbf {\bibinfo
  {volume} {72}},\ \bibinfo {pages} {156} (\bibinfo {year} {1982})}\BibitemShut
  {NoStop}%
\bibitem [{\citenamefont {Kohstall}\ \emph {et~al.}(2011)\citenamefont
  {Kohstall}, \citenamefont {Riedl}, \citenamefont {Guajardo}, \citenamefont
  {Sidorenkov}, \citenamefont {Denschlag},\ and\ \citenamefont
  {Grimm}}]{Kohstall2011}%
  \BibitemOpen
  \bibfield  {author} {\bibinfo {author} {\bibfnamefont {C.}~\bibnamefont
  {Kohstall}}, \bibinfo {author} {\bibfnamefont {S.}~\bibnamefont {Riedl}},
  \bibinfo {author} {\bibfnamefont {E.~R.~S.}\ \bibnamefont {Guajardo}},
  \bibinfo {author} {\bibfnamefont {L.~A.}\ \bibnamefont {Sidorenkov}},
  \bibinfo {author} {\bibfnamefont {J.~H.}\ \bibnamefont {Denschlag}}, \ and\
  \bibinfo {author} {\bibfnamefont {R.}~\bibnamefont {Grimm}},\ }\href
  {\doibase 10.1088/1367-2630/13/6/065027} {\bibfield  {journal} {\bibinfo
  {journal} {New J. Phys.}\ }\textbf {\bibinfo {volume} {13}},\ \bibinfo
  {pages} {065027} (\bibinfo {year} {2011})}\BibitemShut {NoStop}%
\bibitem [{\citenamefont {Chomaz}\ \emph {et~al.}(2015)\citenamefont {Chomaz},
  \citenamefont {Corman}, \citenamefont {Bienaim{\'e}}, \citenamefont
  {Desbuquois}, \citenamefont {Weitenberg}, \citenamefont {Nascimb{\`e}ne},
  \citenamefont {Beugnon},\ and\ \citenamefont {Dalibard}}]{Chomaz:2015eoc}%
  \BibitemOpen
  \bibfield  {author} {\bibinfo {author} {\bibfnamefont {L.}~\bibnamefont
  {Chomaz}}, \bibinfo {author} {\bibfnamefont {L.}~\bibnamefont {Corman}},
  \bibinfo {author} {\bibfnamefont {T.}~\bibnamefont {Bienaim{\'e}}}, \bibinfo
  {author} {\bibfnamefont {R.}~\bibnamefont {Desbuquois}}, \bibinfo {author}
  {\bibfnamefont {C.}~\bibnamefont {Weitenberg}}, \bibinfo {author}
  {\bibfnamefont {S.}~\bibnamefont {Nascimb{\`e}ne}}, \bibinfo {author}
  {\bibfnamefont {J.}~\bibnamefont {Beugnon}}, \ and\ \bibinfo {author}
  {\bibfnamefont {J.}~\bibnamefont {Dalibard}},\ }\href
  {https://www.nature.com/articles/ncomms7162} {\bibfield  {journal} {\bibinfo
  {journal} {Nat. Commun.}\ }\textbf {\bibinfo {volume} {6}},\ \bibinfo {pages}
  {6162} (\bibinfo {year} {2015})}\BibitemShut {NoStop}%
\bibitem [{\citenamefont {Hofferberth}\ \emph {et~al.}(2007)\citenamefont
  {Hofferberth}, \citenamefont {Lesanovsky}, \citenamefont {Fischer},
  \citenamefont {Schumm},\ and\ \citenamefont
  {Schmiedmayer}}]{hofferberth2007nec}%
  \BibitemOpen
  \bibfield  {author} {\bibinfo {author} {\bibfnamefont {S.}~\bibnamefont
  {Hofferberth}}, \bibinfo {author} {\bibfnamefont {I.}~\bibnamefont
  {Lesanovsky}}, \bibinfo {author} {\bibfnamefont {B.}~\bibnamefont {Fischer}},
  \bibinfo {author} {\bibfnamefont {T.}~\bibnamefont {Schumm}}, \ and\ \bibinfo
  {author} {\bibfnamefont {J.}~\bibnamefont {Schmiedmayer}},\ }\href
  {https://www.nature.com/articles/nature06149} {\bibfield  {journal} {\bibinfo
   {journal} {Nature (London)}\ }\textbf {\bibinfo {volume} {449}},\ \bibinfo
  {pages} {324} (\bibinfo {year} {2007})}\BibitemShut {NoStop}%
\bibitem [{foo({\natexlab{c}})}]{footnoteastarvalue}%
  \BibitemOpen
  \href@noop {} {\emph {\bibinfo {title} {\textnormal{The values of $\as^*$
  extracted from experiments and from theory disagree by $4.4\,a_0$. Such a
  mismatch is similar to the one observed in our previous work on the roton
  instability~\cite{Chomaz:2018,Petter:2018}. This behavior might be related to
  the possibility that beyond-mean-field effects are not properly accounted by
  the conventional LHY correction term, see main text. In addition, other
  effects, such as the dynamics and the atom losses~\cite{supmat}, can affect
  the experimental observations.}}}} ({\natexlab{c}})\BibitemShut {NoStop}%
\bibitem [{Not({\natexlab{b}})}]{NoteN}%
  \BibitemOpen
  \href@noop {} {\emph {\bibinfo {title} {\textnormal{This atom number
  corresponds to the condensed atom number measured within the stabilization
  time of $\as$~\cite{supmat}.}}}} ({\natexlab{b}})\BibitemShut {NoStop}%
\bibitem [{\citenamefont {Fisher}\ and\ \citenamefont
  {Ferdinand}(1967)}]{Fisher1967}%
  \BibitemOpen
  \bibfield  {author} {\bibinfo {author} {\bibfnamefont {M.~E.}\ \bibnamefont
  {Fisher}}\ and\ \bibinfo {author} {\bibfnamefont {A.~E.}\ \bibnamefont
  {Ferdinand}},\ }\href {\doibase 10.1103/PhysRevLett.19.169} {\bibfield
  {journal} {\bibinfo  {journal} {Phys. Rev. Lett.}\ }\textbf {\bibinfo
  {volume} {19}},\ \bibinfo {pages} {169} (\bibinfo {year} {1967})}\BibitemShut
  {NoStop}%
\bibitem [{\citenamefont {Imry}\ and\ \citenamefont
  {Bergman}(1971)}]{Imry:1971}%
  \BibitemOpen
  \bibfield  {author} {\bibinfo {author} {\bibfnamefont {Y.}~\bibnamefont
  {Imry}}\ and\ \bibinfo {author} {\bibfnamefont {D.}~\bibnamefont {Bergman}},\
  }\href {\doibase 10.1103/PhysRevA.3.1416} {\bibfield  {journal} {\bibinfo
  {journal} {Phys. Rev. A}\ }\textbf {\bibinfo {volume} {3}},\ \bibinfo {pages}
  {1416} (\bibinfo {year} {1971})}\BibitemShut {NoStop}%
\bibitem [{\citenamefont {Imry}(1980)}]{Imry:1980}%
  \BibitemOpen
  \bibfield  {author} {\bibinfo {author} {\bibfnamefont {Y.}~\bibnamefont
  {Imry}},\ }\href {\doibase 10.1103/PhysRevB.21.2042} {\bibfield  {journal}
  {\bibinfo  {journal} {Phys. Rev. B}\ }\textbf {\bibinfo {volume} {21}},\
  \bibinfo {pages} {2042} (\bibinfo {year} {1980})}\BibitemShut {NoStop}%
\bibitem [{foo({\natexlab{d}})}]{footnoteLuis}%
  \BibitemOpen
  \href@noop {} {\emph {\bibinfo {title} {\textnormal{See the theoretical
  analysis in the revised version of Ref.\,\cite{Tanzi:2018}.}}}}
  ({\natexlab{d}})\BibitemShut {NoStop}%
\bibitem [{\citenamefont {Scarola}\ \emph {et~al.}(2006)\citenamefont
  {Scarola}, \citenamefont {Demler},\ and\ \citenamefont
  {Das~Sarma}}]{Scarola2006}%
  \BibitemOpen
  \bibfield  {author} {\bibinfo {author} {\bibfnamefont {V.~W.}\ \bibnamefont
  {Scarola}}, \bibinfo {author} {\bibfnamefont {E.}~\bibnamefont {Demler}}, \
  and\ \bibinfo {author} {\bibfnamefont {S.}~\bibnamefont {Das~Sarma}},\ }\href
  {\doibase 10.1103/PhysRevA.73.051601} {\bibfield  {journal} {\bibinfo
  {journal} {Phys. Rev. A}\ }\textbf {\bibinfo {volume} {73}},\ \bibinfo
  {pages} {051601} (\bibinfo {year} {2006})}\BibitemShut {NoStop}%
\bibitem [{\citenamefont {Lu}\ \emph {et~al.}(2015{\natexlab{b}})\citenamefont
  {Lu}, \citenamefont {Li}, \citenamefont {Petrov},\ and\ \citenamefont
  {Shlyapnikov}}]{Lu2015}%
  \BibitemOpen
  \bibfield  {author} {\bibinfo {author} {\bibfnamefont {Z.-K.}\ \bibnamefont
  {Lu}}, \bibinfo {author} {\bibfnamefont {Y.}~\bibnamefont {Li}}, \bibinfo
  {author} {\bibfnamefont {D.~S.}\ \bibnamefont {Petrov}}, \ and\ \bibinfo
  {author} {\bibfnamefont {G.~V.}\ \bibnamefont {Shlyapnikov}},\ }\href
  {\doibase 10.1103/PhysRevLett.115.075303} {\bibfield  {journal} {\bibinfo
  {journal} {Phys. Rev. Lett.}\ }\textbf {\bibinfo {volume} {115}},\ \bibinfo
  {pages} {075303} (\bibinfo {year} {2015}{\natexlab{b}})}\BibitemShut
  {NoStop}%
\bibitem [{\citenamefont {Cinti}\ and\ \citenamefont
  {Boninsegni}(2017)}]{Cinti2017}%
  \BibitemOpen
  \bibfield  {author} {\bibinfo {author} {\bibfnamefont {F.}~\bibnamefont
  {Cinti}}\ and\ \bibinfo {author} {\bibfnamefont {M.}~\bibnamefont
  {Boninsegni}},\ }\href {\doibase 10.1103/PhysRevA.96.013627} {\bibfield
  {journal} {\bibinfo  {journal} {Phys. Rev. A}\ }\textbf {\bibinfo {volume}
  {96}},\ \bibinfo {pages} {013627} (\bibinfo {year} {2017})}\BibitemShut
  {NoStop}%
\bibitem [{\citenamefont {Kora}\ and\ \citenamefont
  {Boninsegni}(2019)}]{Youssef2019}%
  \BibitemOpen
  \bibfield  {author} {\bibinfo {author} {\bibfnamefont {Y.}~\bibnamefont
  {Kora}}\ and\ \bibinfo {author} {\bibfnamefont {M.}~\bibnamefont
  {Boninsegni}},\ }\href {https://arxiv.org/abs/1902.08256} {\bibfield
  {journal} {\bibinfo  {journal} {arXiv:1902.08256}\ } (\bibinfo {year}
  {2019})}\BibitemShut {NoStop}%
\bibitem [{\citenamefont {Ronen}\ \emph {et~al.}(2006)\citenamefont {Ronen},
  \citenamefont {Bortolotti},\ and\ \citenamefont {Bohn}}]{Ronen2006bmo}%
  \BibitemOpen
  \bibfield  {author} {\bibinfo {author} {\bibfnamefont {S.}~\bibnamefont
  {Ronen}}, \bibinfo {author} {\bibfnamefont {D.~C.~E.}\ \bibnamefont
  {Bortolotti}}, \ and\ \bibinfo {author} {\bibfnamefont {J.~L.}\ \bibnamefont
  {Bohn}},\ }\href {\doibase 10.1103/PhysRevA.74.013623} {\bibfield  {journal}
  {\bibinfo  {journal} {Phys. Rev. A}\ }\textbf {\bibinfo {volume} {74}},\
  \bibinfo {pages} {013623} (\bibinfo {year} {2006})}\BibitemShut {NoStop}%
\bibitem [{\citenamefont {Lima}\ and\ \citenamefont
  {Pelster}(2011)}]{Pelster:2011}%
  \BibitemOpen
  \bibfield  {author} {\bibinfo {author} {\bibfnamefont {A.~R.~P.}\
  \bibnamefont {Lima}}\ and\ \bibinfo {author} {\bibfnamefont {A.}~\bibnamefont
  {Pelster}},\ }\href {\doibase 10.1103/PhysRevA.84.041604} {\bibfield
  {journal} {\bibinfo  {journal} {Phys. Rev. A}\ }\textbf {\bibinfo {volume}
  {84}},\ \bibinfo {pages} {041604} (\bibinfo {year} {2011})}\BibitemShut
  {NoStop}%
\bibitem [{\citenamefont {Lima}\ and\ \citenamefont
  {Pelster}(2012)}]{Pelster:2012}%
  \BibitemOpen
  \bibfield  {author} {\bibinfo {author} {\bibfnamefont {A.~R.~P.}\
  \bibnamefont {Lima}}\ and\ \bibinfo {author} {\bibfnamefont {A.}~\bibnamefont
  {Pelster}},\ }\href {\doibase 10.1103/PhysRevA.86.063609} {\bibfield
  {journal} {\bibinfo  {journal} {Phys. Rev. A}\ }\textbf {\bibinfo {volume}
  {86}},\ \bibinfo {pages} {063609} (\bibinfo {year} {2012})}\BibitemShut
  {NoStop}%
\bibitem [{\citenamefont {Cikojević}\ \emph {et~al.}(2018)\citenamefont
  {Cikojević}, \citenamefont {Markić}, \citenamefont {Astrakharchik},\ and\
  \citenamefont {Boronat}}]{Astrakharchik:2018}%
  \BibitemOpen
  \bibfield  {author} {\bibinfo {author} {\bibfnamefont {V.}~\bibnamefont
  {Cikojević}}, \bibinfo {author} {\bibfnamefont {L.~V.}\ \bibnamefont
  {Markić}}, \bibinfo {author} {\bibfnamefont {G.~E.}\ \bibnamefont
  {Astrakharchik}}, \ and\ \bibinfo {author} {\bibfnamefont {J.}~\bibnamefont
  {Boronat}},\ }\href {https://arxiv.org/abs/1811.04436} {\bibfield  {journal}
  {\bibinfo  {journal} {arXiv:1811.04436}\ } (\bibinfo {year}
  {2018})}\BibitemShut {NoStop}%
\bibitem [{\citenamefont {Barone}\ and\ \citenamefont
  {Paternò}(1982)}]{Barone:101094}%
  \BibitemOpen
  \bibfield  {author} {\bibinfo {author} {\bibfnamefont {A.}~\bibnamefont
  {Barone}}\ and\ \bibinfo {author} {\bibfnamefont {G.}~\bibnamefont
  {Paternò}},\ }\href {http://cds.cern.ch/record/101094} {\emph {\bibinfo
  {title} {Physics and applications of the Josephson effect}}}\ (\bibinfo
  {publisher} {Wiley},\ \bibinfo {address} {New York, NY},\ \bibinfo {year}
  {1982})\BibitemShut {NoStop}%
\bibitem [{foo({\natexlab{e}})}]{footnotecentering}%
  \BibitemOpen
  \href@noop {} {\emph {\bibinfo {title} {\textnormal{We note that we have also
  checked our analysis without performing the recentering step and the same
  features remain. For instance, for our test data of
  Fig.\,\ref{fig:coherence}, the effect being mainly that the side peaks in (e)
  are more washed out and a slight difference occurs between $n_{\mathcal{M}}$
  and $n_\Phi$, both showing still side peaks.}}}} ({\natexlab{e}})\BibitemShut
  {NoStop}%
\bibitem [{\citenamefont {Ilzh\"ofer}\ \emph {et~al.}(2018)\citenamefont
  {Ilzh\"ofer}, \citenamefont {Durastante}, \citenamefont {Patscheider},
  \citenamefont {Trautmann}, \citenamefont {Mark},\ and\ \citenamefont
  {Ferlaino}}]{Ilzhoefer2018}%
  \BibitemOpen
  \bibfield  {author} {\bibinfo {author} {\bibfnamefont {P.}~\bibnamefont
  {Ilzh\"ofer}}, \bibinfo {author} {\bibfnamefont {G.}~\bibnamefont
  {Durastante}}, \bibinfo {author} {\bibfnamefont {A.}~\bibnamefont
  {Patscheider}}, \bibinfo {author} {\bibfnamefont {A.}~\bibnamefont
  {Trautmann}}, \bibinfo {author} {\bibfnamefont {M.~J.}\ \bibnamefont {Mark}},
  \ and\ \bibinfo {author} {\bibfnamefont {F.}~\bibnamefont {Ferlaino}},\
  }\href {\doibase 10.1103/PhysRevA.97.023633} {\bibfield  {journal} {\bibinfo
  {journal} {Phys. Rev. A}\ }\textbf {\bibinfo {volume} {97}},\ \bibinfo
  {pages} {023633} (\bibinfo {year} {2018})}\BibitemShut {NoStop}%
\end{thebibliography}%

\appendix
\renewcommand\thefigure{\thesection S\arabic{figure}}   
\setcounter{figure}{0}   
\section{Supplemental Material}

\section*{Ground state calculations}

We perform numerical calculations of the ground state following the procedure detailed in the supplementary information of Ref.\,\cite{Chomaz:2018}. The calculations are based on the conjugate-gradients technique to minimize the energy functional of an eGPE~\cite{Ronen2006bmo}. 
In particular, the eGPE accounts for the effect of quantum fluctuations, by including the LHY term $\Delta \mu[n] = 32 g (n \as)^{3/2}(1 + 3 \edd^2 / 2) / 3 \sqrt{\pi}$ in the system's Hamiltonian (here $g = 4\pi \hbar^2 \as/m$ and $n=|\psi|^2$ is the spatial density of the macroscopic state $\psi$). $\Delta \mu[n]$ has been obtained under a local density approximation in Refs.\,~\cite{Pelster:2011,Pelster:2012}. The relevance of the LHY correction has been demonstrated in various studies of dipolar Bose gases close to the mean-field instability~\cite{Waechtler:2016,Waechtler:2016b,Bisset:2016, Chomaz:2016, Schmitt:2016, Chomaz:2018} as it brings an additional repulsive potential, stabilizing the gas against mean-field collapse at large density. We note that the exact functional form of the potential, originating from beyond mean-field effects, has been questioned by several experimental results in finite-size trapped systems~\cite{Schmitt:2016,Chomaz:2018,Igor:2018, Cabrera2018}, calling for further theory developments~\cite{Astrakharchik:2018}.

Our numerical calculations provide us with the three-dimensional ground-state wavefunctions $\psi(\boldsymbol{r})$. From this, we compute the axial in-situ density profile along the trap's weak axis, $n(y)=\int |\psi(\bs r)|^2 dx dz$ and find density profiles, corresponding to the BEC, the supersolid or the ID phase, that we plot in Fig.\,\ref{fig:theory}.
From the density profiles that exhibit a density modulation, we evaluate $S$ by performing Gaussian fits to each droplet, i.\,e.\,to $n(y)$ with $y$ ranging between two neighboring local density minima. From these Gaussian fits, we evaluate the sets of centers $\{y^{(0)}_i\}_i$ and widths $\{\sigma_i\}_i$ corresponding to the macroscopic Gaussian wavefunctions $\{\psi_i\}_i$ associated to the individual droplets in the array. We then approximate the droplet wavefunction via $\psi_i(y)\approx\sqrt{n(y\approx y^{(0)}_i)}= \alpha_i \exp \left(-(y-y^{(0)}_i)^2/2\sigma_i^2\right)$ with $\alpha_i$ a normalization coefficient such that $\int |\psi_i(y)|^2 dy=1$. We then evaluate the wavefunction overlap $S_i$ between the neighboring droplets $i-1$ and $i$ via:
\begin{eqnarray}
    S_i &\equiv& \int \psi_{i-1}^*(y)\psi_i(y)dy \\
    \label{eq:Si}
    &=& \sqrt{\frac{2\sigma_i\sigma_{i-1}}{\sigma_i^2+\sigma_{i-1}^2}} \exp\left(-\frac{(y^{(0)}_i-y^{(0)}_{i-1})^2}{2(\sigma_i^2+\sigma_{i-1}^2)}\right).
\end{eqnarray}
The latter equation is obtained via an analytical evaluation of the Gaussian integral. The characteristic link strength defined in the paper is then computed by averaging $S_i$ over all droplet links in the array: $S=\langle S_i \rangle_i$. In our calculation, we only consider as droplets all density peaks of at least 10\,\% of the global density maximum.

\section*{Link Strength and estimate of tunneling rate}

Generally speaking, the wavefunction overlap between neighboring droplets relates to a tunneling term, which sets a particle exchange term between two neighboring droplets~\cite{Josephson:1962,Barone:101094,Javanainen:1986,Smerzi}. Following the work of Ref.\,\cite{Wenzel:2018}, we perform a first estimate of the tunneling coefficient by simply considering the single-particle part of the Hamiltonian and evaluate it between two neighboring droplets. We note that, in our particular setting where the density modulation is not externally imposed  but arises from the mere interparticle interactions, the inter-droplet interaction may also play a crucial role. To perform a more precise estimation of the tunneling between droplets, one would certainly need to properly account for this effect. Here, we stress that our approach simply gives a rough idea of the magnitude of tunneling while it does not aim to be a quantitative description of it. This consideration calls for further studies making a systematic analysis of the full Hamiltonian and of the full phase diagram within the Josephson junction formalism and beyond.

Generalizing the description of Ref.\,\cite{Wenzel:2018} to neighboring droplets of different sizes and amplitudes, which are described by a three-dimensional wavefunction $\psi_i(\bs r)$ approximated to a three-dimensional Gaussian of widths $(\sigma_{i,x}, \sigma_{i,y}, \sigma_{i,z})$ with $ \sigma_{i,y}=\sigma_i$, our estimate writes: 
\begin{eqnarray}
    \label{eq:Ji}
    J_i &=& \frac{\hbar^2 S_i}{2m}\left[\sum_{k=x,y,z}\frac{1+\left(\frac{\sigma_{i,k}\sigma_{i-1,k}}{\ell_k^2}\right)^2}{\sigma_{i,k}^2+\sigma_{i-1,k}^2}\right.\nonumber\\
    &&+\left.\frac{(y^{(0)}_i-y^{(0)}_{i-1})^2}{2\sigma_{i}\sigma_{i-1}}\frac{\left({\sigma_{i}\sigma_{i-1}}/{\ell_y}\right)^4-1}{\sigma_{i}^2+\sigma_{i-1}^2}\right],
\end{eqnarray}
 where $\ell_{x,y,z}=\sqrt{\hbar/m\omega_{x,y,z}}$ are the harmonic oscillator lengths. 
 
 In general, the tunnelling coefficients set two typical rates relevant for equilibration processes. The first one is the bare single-particle tunneling rate, which is equal to $J_i/h$, while the second accounts for the bosonic enhancement from the occupation of the droplet modes and writes  $\tilde{t}_i=\sqrt{N_iN_{i-1}}|J_i|/h$ where $N_i$ is the number of atoms in droplet $i$. In our analysis, we then define the average rates over the droplet arrays as  characteristic rates $J/h=\langle J_i \rangle_i/h $, and $\tilde{t}=\langle \tilde{t}_i \rangle_i$; see e.g.~\cite{Hadzibabic}. While the ground state evolves from a BEC to a supersolid to an ID, the relevant timescale for achieving (global) equilibrium crosses from being set by the trap frequencies to the above-mentioned tunneling rates. 

Using our approximate model, we here give a first estimate of the rates $J/h$ and $\tilde{t}$ as a function of $\as$, for the parameters of Fig.\,\ref{fig:theory}(b-d) of the main text (i.e. Er quantum gas with $N=5\times 10^4$ atoms). Here we find that, for $\as=\as^*$, $J/h\sim 400\,$Hz and $\tilde{t} \sim 10\,$MHz while for $\as=\as^*-2.5\,a_0$, $J/h\sim 10^{-7}\,$Hz and $\tilde{t} \sim 10^{-3}\,$Hz.   



\section*{Toy model for the interference pattern}

As described in the main text we use a simple toy model, adapted from Ref.\,\cite{Hadzibabic}, to identify the main features of the TOF interference patterns obtained from an insitu density-modulated state. As a quick reminder, our model considers a one-dimensional array of $N_D$ Gaussian droplets, described by a single classical field, $\psi_i$, thus neglecting quantum and thermal fluctuations. We compute the TOF density distribution from the free-expansion of the individual $\psi_i$ during a time $t$ via $n(y,t) = |\sum_i \psi_i(y,t)|^2$. In our calculations, we also account for the finite imaging resolution by convolving the resulting $n(y,t)$ with a gaussian function of width $\sigma_{\rm im}$. Here we allow the characteristics of the individual $\psi_i$ to fluctuate. In this aim, we introduce noise on the corresponding parameter with a normal distribution around its expectation value and with a variable standard deviation (only $\phi_i$ can also have a uniform distribution). We then perform a Monte-Carlo study and perform ensemble averages, similar to our experimental analysis procedure. We note that, in this simple implementation, the noise on the different parameters -- droplet amplitudes, widths and distances -- are uncorrelated.

In the main text, we present results for a single set of parameters, namely $N_D=4$, $d\equiv\langle d_i\rangle_i=2.8\,\um$ (mean droplet distance),  $\sigma_y\equiv\langle \sigma_i\rangle_i=0.56\,\um$ (mean droplet size), $t=30\,$ms, and $\sigma_{\rm im} = 3\,\um$, typical for our experimental Er setting and the corresponding theory expectations in the supersolid regime. $\langle \cdot \rangle_i$ denotes the average over the droplets. In  this section, we have a deeper look at the impact of the different parameters on both the TOF signal and our FT analysis. We study both the fully phase coherent and fully incoherent case, and the unchanged parameters are set as in Fig.\,\ref{fig:coherence}(j,m) and (l,o).

\begin{figure}[htbp!]
	\includegraphics[width=0.5\textwidth]{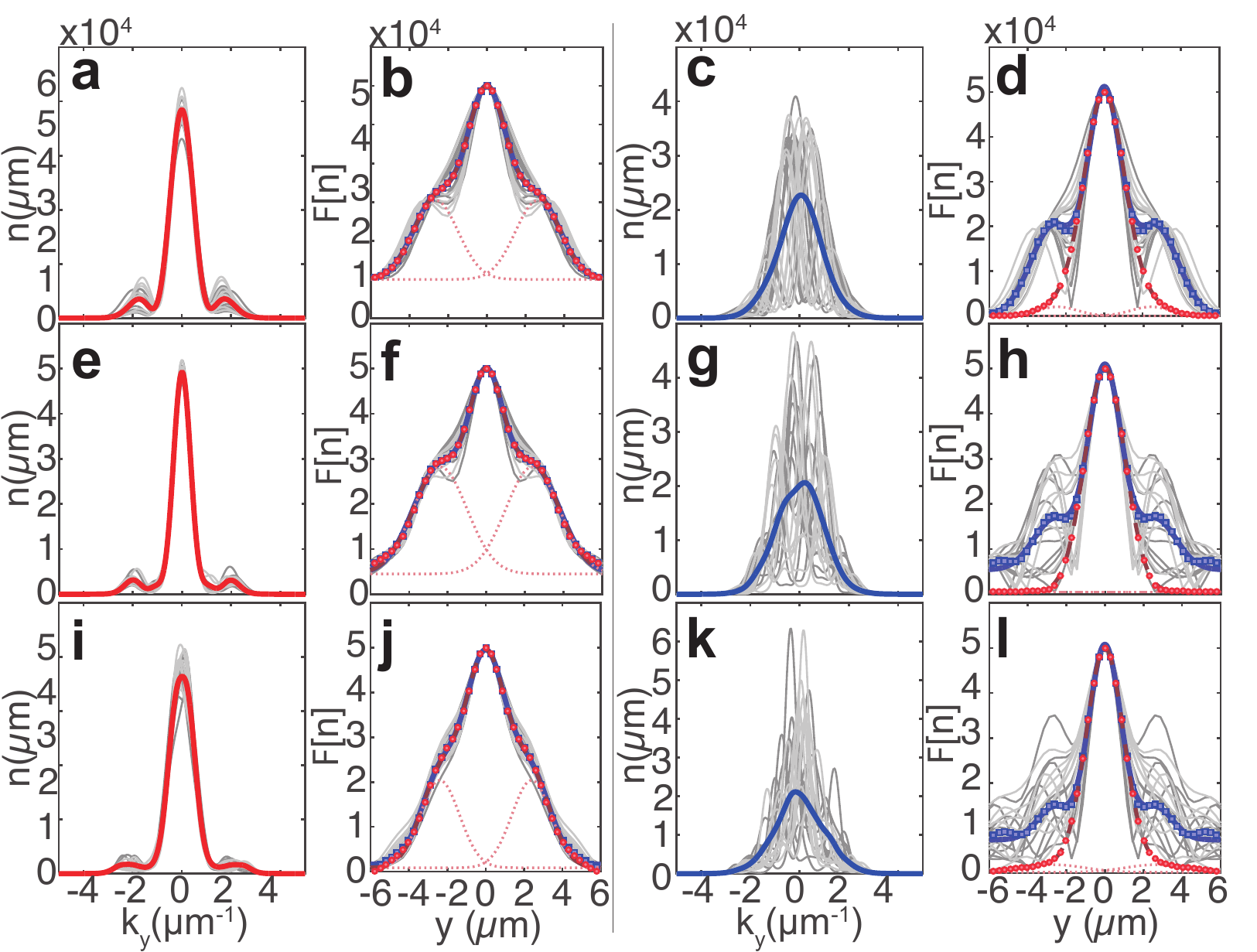}
	\caption{\label{fig:toymodel_ND} \textbf{Toy model realizations with varying number of droplets $N_D$}. We use 100 independent draws, and expectation values $d=2.85\,\um$,  $\sigma_y=0.56\,\um$ (with 10\% noise) and either $\phi_i=0$ (a,b,e,f,i,j), or $\phi_i$ uniformly distributed between 0 and $2\pi$ (c,d,g,h,k,l). (a--d)~$N_D=2$, (e--h)~$N_D=3$ and (i--l) $N_D=8$. (a,c,e,g,i,k) TOF density profiles and (b,d,f,h,j,l) corresponding FT analysis of the interference patterns, same color code as Fig\,.\ref{fig:coherence}.  
	}
\end{figure}

In Fig.\,\ref{fig:toymodel_ND}, we first exemplify the TOF and FT profiles for a varying number of droplets, between 2 and 8, which cover the range of relevant $N_D$ over the phase diagram of Fig.\,\ref{fig:theory}. The results remain remarkably similar to the realization of Fig.\,\ref{fig:coherence} with only slight quantitative changes. The main difference lies in the individual interference patterns obtained in the phase incoherent case. With increasing $N_D$, those profiles become more complex and made of a larger number of peaks (see (c,g,k)). Yet, in this incoherent case, a similar (non-modulated) profile is recovered in the averaged $n(k_y)$ for all $N_D$. Additionally, we note that for the coherent case with $N_D=8$, the side peaks in the FT analysis (see (j)) become less visible. By performing additional tests, we attribute this behavior to the limited TOF duration, $t$, used in our experiment yielding a typical length scale, $\sqrt{\hbar t/m}$ ($= 3.4 \um$), which becomes small compared to the system size ($\approx (N_D-1)d+\sigma_y$) for large $N_D$. This intermediate regime in the TOF expansion leads to more complex features, including smaller-sized motifs, in the interference patterns. Finally, when accounting for our imaging resolution, it yields a broadening of the structure observed in the TOF images and less visible peaks in the FT (see (i)). We note that our experiments, because of limited $N$ and additional losses, should rather lie in the regime $2 \leq N_D\leq 5$; see Fig.\,\ref{fig:theory}(b).

\begin{figure}[htbp!]
	\includegraphics[width=0.5\textwidth]{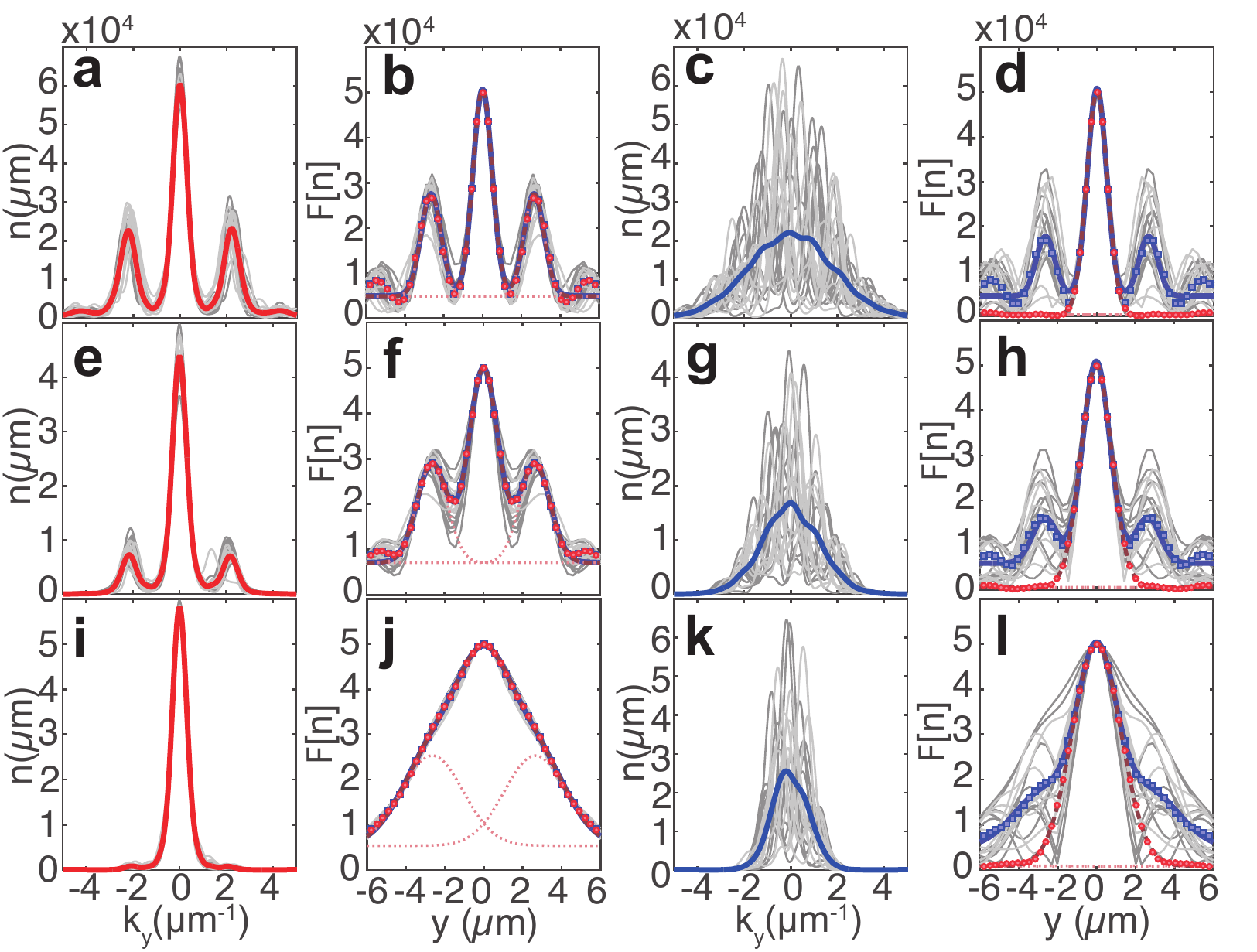}
	\caption{\label{fig:toymodel_elld} \textbf{Toy model realizations with varying $\sigma_y/d$}. We use 100 independent draws, with $N_D=4$, $d=2.85\,\um$ (with 10\% noise) and either $\phi_i=0$ (a,b,e,f,i,j), or $\phi_i$ uniformly distributed between 0 and $2\pi$ (c,d,g,h,k,l). For each realization we also compute the associated mean $S$. (a--d) $\sigma_y/d=0.1$, yielding $S=1.8\times 10^{-7}$ (e--h)$\sigma_y/d=0.15$, matching $S=1.7\times 10^{-4}$ and (i--l) $\sigma_y/d=0.25$, matching $S=0.028$. (a,c,e,g,i,k) TOF density profiles and (b,d,f,h,j,l) Corresponding FT analysis of the interference patterns, same color code as Fig.\,\ref{fig:coherence}. 
	}
\end{figure}

We then investigate the evolution of the interference patterns and their FT analysis for a varying mean droplet size, $\sigma_y$, while keeping their mean distance, $d$, fixed. This study is particularly relevant recalling that, within the Josephson junction formalism (see main text and corresponding section of this Supplemental Material), the key parameter controlling the tunneling rate between the droplets is set by the ratio $\sigma_y/d$, and the link strength parameter that we use to characterize the supersolid regime scales roughly as $\exp(-(d/2\sigma_y)^2)$. Thus, in our experiment, $\sigma_y/d$ is intrinsically expected to decrease with the scattering length (see Fig.\,\ref{fig:ContrastVSa}). Performing a direct estimate of the average droplet link from the initial state of our toy model, we find $S=0.004$ for the calculations of Fig.\,\ref{fig:coherence}(j-o), lying in an expected supersolid regime yet rather close to the supersolid-to-ID transition. 
Figure \ref{fig:toymodel_elld} investigates the effect of smaller and larger values of $\sigma_y/d$ (and consequently of $S$) on the TOF and FT profiles while independently assuming phase coherence or incoherence.  
Qualitatively, the features remain similar as in Fig.\,\ref{fig:coherence}(j-o). In the coherent case, side peaks are visible in the individual as well as in the mean $n(k_y)$ (see (a,e,i)) and yield side peaks in the FT profiles, with $n_\mathcal{M}\approx n_\mathcal{\Phi}$ (see (b,f,j)). Increasing (decreasing) $\sigma/d$ mainly results in a stronger (weaker) signal both in the interference pattern and their FT analysis. Within our toy model, we find that, already for $\sigma/d=0.25$, the signal nearly vanishes; see (i,j). Even if, given the approximations used in our toy model, this exact value may not fully hold for our experimental conditions, we expect a similar trend. It is interesting to keep in mind that this effect may limit our capacity of detecting an underlying supersolid state via matter-wave interference in experiments. In the incoherent case, the effect of decreasing $\sigma_y/d$ mainly results in a broader shape of the mean density profile, while it remains non-modulated; see (c,g,k). In the FT analysis $n_{\Phi}$ remains structure-less independently of $\sigma_y/d$ while the structures in $n_\mathcal{M}$ becomes sharper with decreasing  $\sigma_y/d$, as in the coherent case; see (d,h,l).

\begin{figure}[htbp!]
	\includegraphics[width=0.5\textwidth]{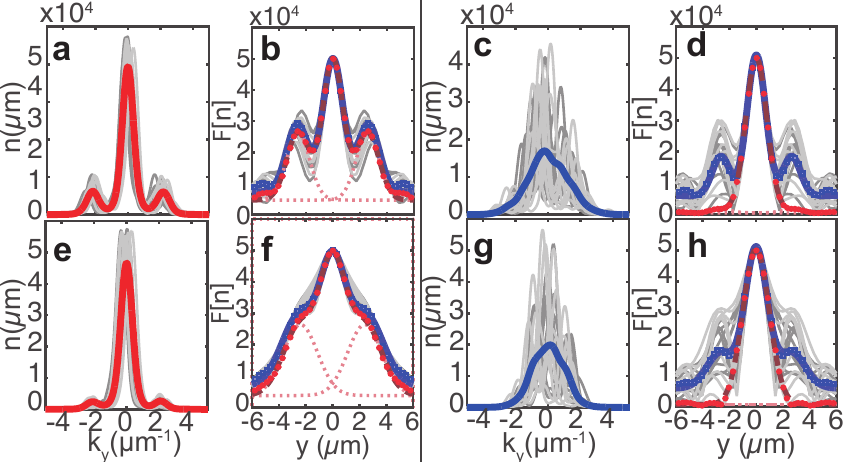}
	\caption{\label{fig:toymodel_center} \textbf{Toy model realizations allowing noise in the center position}. We use 100 independent draws, with $N_D=4$, $d=2.85\,\um$ (with 10\% noise), $\sigma_y/d=0.15$ (a--d) or $\sigma_y/d=0.2$ (e--h),  and either $\phi_i=0$ (a,b,e,f,i,j), or $\phi_i$ uniformly distributed between 0 and $2\pi$ (c,d,g,h,k,l). Center fluctuation are introduced as normal noise around 0 with standard deviation of $2\,\um^{-1}$ in situ (a,c,e,g,i,k) TOF density profiles and (b,d,f,h,j,l) corresponding FT analysis of the interference patterns, same color code as Fig.\,\ref{fig:coherence}.
	}
\end{figure}

Finally, we investigate how a possible shot-to-shot noise on the position of the central interference peak could affect our observables of the density modulation and phase coherence. In the experiments, such fluctuations may occur, for instance, because of beam-pointing fluctuations or excitations of the gas. Although we compensate for such effects by recentering the individual images (see Imaging Analysis section), residual effects 
may remain, in particular due to center misestimation in the mere presence of the interference patterns of interest. To investigate this aspect, we repeat our toy model calculations now including noise in the  global droplet array position and using a standard deviation of $2\,\um$ for two values of $\sigma_y/d$; see Fig.\ref{fig:toymodel_center}. Again, qualitatively the observed features remains similar to our prediction in the main text. The main effect lies in the appearance of a small discrepancy in the coherent case between $n_{\Phi}$ and $n_{\mathcal{M}}$, while the structure in the incoherent case remains similar. As the center misestimation should be the most severe in the latter case (due to the variability of the interference patterns observed here), our test shows the robustness of our analysis procedure against this issue.

\section*{Imaging Analysis: $\Dy$ and $\Er$}
The density distributions in momentum space are extracted from the TOF images using the free-expansion expectation. In the Dy case, the thermal component is subtracted from the individual distribution by cutting out the central region of the cloud and performing an isotropic Gaussian fit on the outer region. This subtraction is beneficial because of the large thermal fraction. In the $\Er$ case, such a subtraction is on the contrary complicated because of the weak thermal component and this pre-treatment may lead to improper estimation of $A_\mathcal{M}$ and $A_\Phi$ in the later analysis. The obtained momentum density distributions are then recentered and integrated numerically along $k_z$($k_x$) between $[-2.0,+2.0]\,\mu$m$^{-1}$ ($[-1.28,+1.28]\,\mu$m$^{-1}$) to obtain $n(k_Y)$ ($n(k_y)$) for $\Dy$ ($\Er$). The recentering procedure uses the result a single Gauss fit to the TOF images. The fit is performed after convoluting each image with a Gaussian function of width $0.5\,\um$ whose purpose is to reduce the impact of the interference pattern on the center estimation~\cite{footnotecentering}. 

In order to characterise the system's state, we use the Fourier transform, $\mathcal{F}[n](y)$ of the single density profile, $n(k_y)$. We then compute two average profiles,  $n_\mathcal{M}$ and $n_\Phi$,  relying on ensemble average over all measurements under the same experimental conditions; see below for a detailed discussion on $n_\mathcal{M}$ and $n_\Phi$.  In all the measurements reported in this work we use averages over typically 15 to 100 realizations. 

To quantify both the existence of a density modulation and global phase coherence on top of this modulation, we fit both $n_\mathcal{M}(y)$ and $n_\Phi(y)$ with a triple-Gaussian function, where one Gaussian accounts for the central peak and the other Gaussians are accounting for the symmetric side peaks. The amplitudes of the latter give $A_{\mathcal{M}}$ and $A_{\Phi}$, respectively. The distance between the side peaks and the central one is allowed to vary between $[2.5, 2.7]\,\mu$m ($[2.3,2.5]\,\mu$m) in the case of $\Dy$ ($\Er$).

\section*{Details on the Fourier analysis}

In our analysis we rely on two averaged profiles, named $n_\mathcal{M}$ or $n_\Phi$, to quantify both the density modulation and its phase coherence. Here we detail the meaning of the average performed.

The Fourier transform (FT) of the integrated momentum distributions, $n(k_y)$, which reads $\mathcal{F}[n](y)=\left|\mathcal{F}[n](y)\right|\,\exp(i\arg\left(\mathcal{F}[n](y)\right))$ sets the ground for our analysis. 
As stated in the main text, an in-situ density modulation of wavelength $y^*$ yields patterns in $n(k_y)$ and consequently induce peaks at $y\approx y^*$, in the FT norm, $\left|\mathcal{F}[n](y)\right|$, see Fig.\,\ref{fig:coherence}(g-i) and (m-o). Spatial variations of the phase relation within the above-mentioned density modulation translate into phase shifts of the interference patterns, which are stored in  the FT argument at $y\approx y^*$, $\arg\left(\mathcal{F}[n](y^*)\right)$; see also Ref.\,\cite{Hadzibabic,hofferberth2007nec}.

The first average that we use is  $n_\mathcal{M}(y)= \langle |\mathcal{F}[n](y)|\rangle$, i.\,e.\ the  average of the FT norm of the individual images. As the phase information contained in $\arg\left(\mathcal{F}[n](y)\right)$ is discarded from $n_{\mathcal{M}}$ when taking the norm,  the peaks in $n_{\mathcal{M}}$ probe the mere existence 
of an insitu density modulation of roughly constant spacing within the different realizations. The second average of interest is $n_\Phi(y)=|\langle \mathcal{F}[n](y)\rangle|$, i.\,e.\,the average of the full FT of the individual images. In contrast to $n_\mathcal{M}$, $n_\Phi$ keeps the phase information of the individual realizations contained in $\arg\left(\mathcal{F}[n](y^*)\right)$. Consequently, peaks in $n_\Phi$ indicate that the phase relation is maintained over the density modulation, in a similar way for all realizations. Their presence thus provides information on the global phase coherence of a density-modulated state.

\section*{Experimental sequence: $\Dy$ and $\Er$}

\textit{$^{166}$Erbium -}
The BEC of $\Er$  is prepared similarly to Refs.\,\cite{Aikawa:2012,Chomaz:2016,Chomaz:2018,Petter:2018}. We start from a magneto-optical trap with $2.4\times10^7$ \Er\ atoms at a temperature of $10\mu{\rm K}$, spin-polarized in the lowest Zeeman sub-level. In a next step we load about $3\times10^6$ atoms into a crossed optical dipole trap (ODT) operated at 1064\,nm.
We evaporatively cool the atomic cloud by reducing the power and then increasing the ellipticity of one of the ODT beams. During the whole evaporation a constant magnetic field of $B=1.9\,$G ($\as=80\,a_0$) along $z$ is applied. We typically achieve BEC with $1.4\times10^5$ atoms and a condensed fraction of $70\%$. In a next step the ODT is reshaped in 300\,ms into the final trapping frequencies $\omega_{x,y,z}=2\pi\times(145,31.5,151)\,$Hz. Consecutively, we ramp $B$ linearly to $0.62\,$G ($64.5\,a_0$) in 50\,ms and obtain a BEC with $8.5\times10^4$ atoms, which are surrounded by $3.5\times10^4$ thermal atoms. This point marks the start of the ramp to the final $a_s$.

\textit{$^{164}$Dysprosium -}
For the production of a $^{164}\text{Dy}$ BEC we closely follow the scheme presented in \cite{Trautmann2018}. Starting from a $3\,$s loading phase of our 5-beam MOT in open-top configuration \cite{Ilzhoefer2018}, we overlap a $1064\,$nm single-beam dipole trap with a $\nicefrac{1}{e^2}$-waist of about $22\,\mu$m, for $120\,$ms. Eventually, we transfer typically $8\times10^6$ atoms utilizing a time averaging potential technique to increase the spatial overlap with the MOT. After an initial $1.1\,$s evaporative cooling phase by lowering the power of the beam, we add a vertically propagating beam, derived from the same laser, with a $\nicefrac{1}{e^2}$-waist of about $130\,\mu$m to form a crossed optical dipole trap for additional confinement. Subsequently, we proceed forced evaporative cooling to reach quantum degeneracy by nearly exponentially decreasing the laser powers in the two dipole-trap beams over $3.6\,$s.
We achieve BECs of $^{164}\text{Dy}$ with typically $10^5$ atoms and condensate fractions of about $40\%$. During the entire evaporation sequence the magnetic field is kept constant at $2.5\,$G pointing along the vertical ($z$-) axis.\\

To be able to condense directly into the supersolid, we modify the dipole trap to condense at a stronger confinement of $\omega_{x,y,z}=2\pi\times\left(225,37,134\right)\,$Hz. After a total evaporative cooling duration of $3.1\,$s, we achieve Bose-Einstein condensation at $2.55\,$G and reach a state with supersolid properties at $2.43\,$G, keeping the magnetic field constant throughout the entire evaporation sequence for both cases.\\

\textit{Time of flight and imaging for $\Er$ and $\Dy$ -}
In order to probe the momentum distribution of the Dy (Er) gases, we switch off the confining laser beams and let the atoms expand freely for $18\,$ms (15\,ms), while keeping the magnetic field constant. Consecutively the amplitude of $B$ is increased to a fixed amplitude of 5.4\,G (0.6\,G). In the case of $\Dy$, the magnetic field orientation is rotated in order to point along the imaging axis. This ensures constant imaging conditions for different $\as$. After an additional $9\,$ms (15\,ms) we perform a standard absorption imaging.

\section*{Tuning the scattering length in $\Er$ and $\Dy$}

\textit{$^{166}$Erbium -} All measurements start with a BEC at $64.5\,a_0$. In order to probe the BEC-supersolid-ID region, we linearly ramp $a_s$ to its target value in $\tr = 20\,$ms by performing a corresponding ramp in $B$. Due to a finite time delay of the magnetic field in our experimental setup and the highly precise values of $a_s$ needed for the experiment, we let the magnetic field stabilize for another 15\,ms before $t_h=0\,$ starts. By this, we ensure that the residual lowering of $a_s$ during the entire hold time is $\lesssim0.3\,a_0$. In the main text, we always give the $a_s$ at $\tho=0\,$. Furthermore, we estimate our magnetic field uncertainty to be $\pm2.5\,$mG, leading to a $\pm0.2\,a_0$ uncertainty of $a_s$ in our experiments.

\begin{figure}[htbp!]
	\includegraphics[width=1\linewidth]{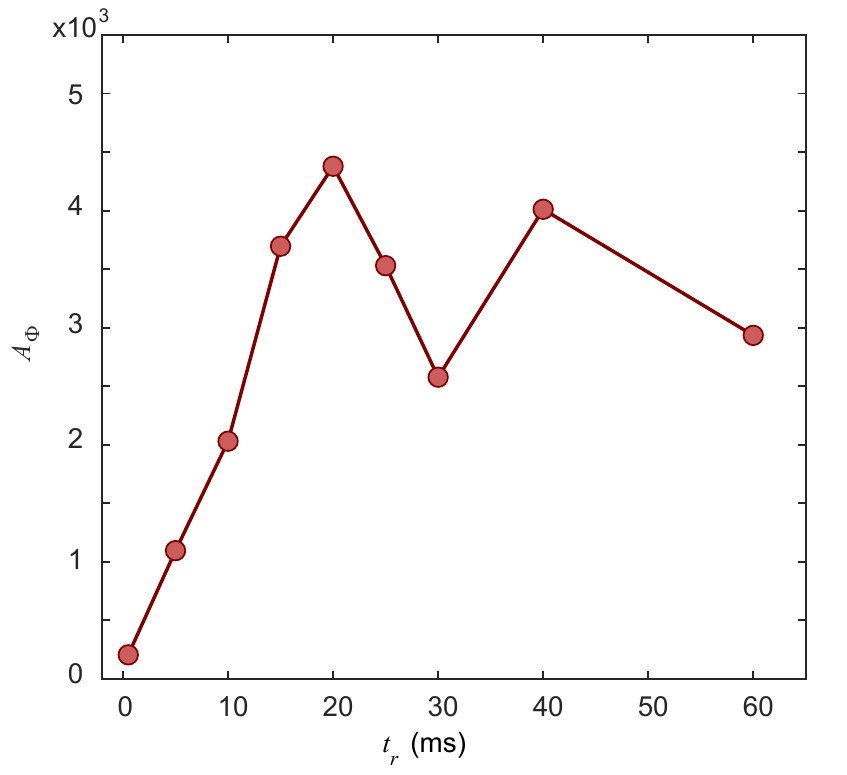}
	\caption{\label{fig:VarRampTimeEr}
	\textbf{Ramp time effect on the supersolid behavior}
	Measured $A_\Phi$ for various durations of the scattering-length ramp with $\Er$ and a final $\as=54.1(2)\,a_0$. All measurements include a 15\,ms stabilization time after $\tr$ and are performed with an additional hold of $\tho=5\,$ms.  
	}
\end{figure}

To choose the best ramping scheme, we have performed experiments varying $\tr$ from 0.5\,ms to 60\,ms, ramping to a fixed $\as$ lying in the supersolid regime, and holding for $\tho=5\,$ms after a fixed 15\,ms waiting time. We record the evolution of $A_\Phi$ as a function of $\tr$; see Fig.\,\ref{fig:VarRampTimeEr}. When increasing $\tr$, we first observe that $A_\Phi$ increases, up to $\tr=20$\,ms, and then $A_\Phi$ gradually decreases. The initial increase can be due to diabatic effects and larger excitation when fast-crossing the phase transition. 
On the other hand, the slow decrease at longer $\tr$ can be explained by larger atom loss during the ramp. We then choose $\tr=20\,$ms as an optimum value where a supersolid behavior develops and maintains itself over a significant time  while the losses are minimal.

\begin{figure}[htbp!]
	\includegraphics[width=\linewidth]{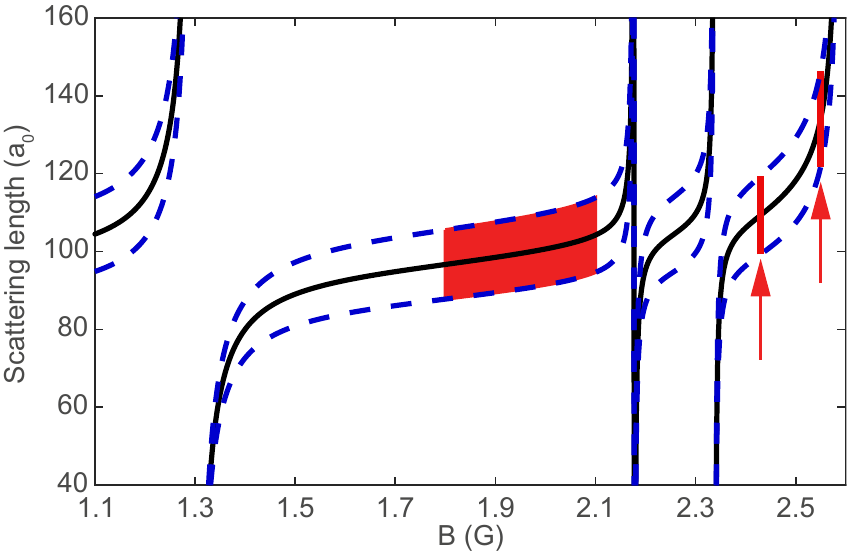}
	\caption{\label{fig:Dy_as_vs_B} \textbf{Estimated scattering length tuning in $\Dy$}
	Estimated dependence of $\as$ on $B$ for $\Dy$. The FR positions and widths have been extracted from trap-loss spectroscopy measurements, the background scattering length is estimated	to $a_{\rm bg}=87(8)\,a_0$, see text. The blue dashed line gives an error-estimate considering only the errorbar on $a_{\rm bg}$ from the mere $\as$ measurement of Ref.~\cite{Tang2015} and not accounting for uncertainty of the Feshhach scan.  For Figs.\,\ref{fig:DecayRateCoherenceA}-\ref{fig:DecayRateCoherence}, we use $B$ between 1.8\,G and 2.1\,G (red area); for Figs.\,\ref{fig:Evap}--\ref{fig:EvapLife}, we keep at two constant $B$-values, namely  2.43\,G and 2.55\,G (red arrows).
	}
\end{figure}

\textit{$^{164}$Dysprosium -} As the value of the background scattering, $a_{\rm bg}$ length for \Dy is still under debate~\cite{Tang2015,Schmitt:2016,Igor:2018}, we discuss the experimental settings in terms of magnetic field. Yet, to gain a better understanding of the tunability of $\as$ in our experiment, we first perform a Feshbach spectroscopy scan on a BEC at $T=60\,$nK. After evaporative cooling at $B=2.5\,$G, we jump to $B$ varying from $1\,$G to $7.5\,$G and we hold the sample for $100\,$ms. Finally, we switch off the trap, let the cloud expand for $26\,ms$ and record the total atom number as a function of $B$. We then fit the observed loss features with a gaussian fit to obtain the position $B_{0,i}$ and width $\Delta B_{i}$ of the FRs, numbered ${i}$. We finally use the standard Feshbach resonance formula to estimate the $\as$-to-$B$ dependence via $\as(B)=a_{\rm bg}\prod_i\left(1-\Delta B_{i}/(B-B_{0,i})\right)$. Here we account for 8 FRs located between 1.2\,G and 7.2\,G. Depending on the background scattering length $a_{\rm bg}$, the overall magnitude of $\as(B)$ changes. We can get an estimate of  $a_{\rm bg}$ from literature. In Fig.\,\ref{fig:Dy_as_vs_B}, we use the value of $\as$ from Ref.~\cite{Tang2015} obtained at $1.58$\,G close to the $B$-region investigated in our experiment, $\as=92(8)\,a_0$. By reverting the $\as(B)$ formula, we set $a_{\rm bg}=87(8)\,a_0$. 
For the measurements of Figs.\,\ref{fig:DecayRateCoherenceA}-\ref{fig:DecayRateCoherence}, we ramp $B$ linearly from 2.5\,G in 20\,ms to a final value ranging from 1.8 to 2.1\,G, for which we estimate $\as$ ranging from $97(9)\,a_0$ to $105(10)\,a_0$. We calibrate our magnetic field using RF spectroscopy, with a stability of about 2\,mG. In the Dy case, we do not apply an additional stabilization time. This is justified because of the more mellow $\as$-to-$B$ dependence in the $B$-range of interest as well as of the wider $\as$-range of the superoslid regime (see Fig.\,\ref{fig:theory}) compared to the Er case.
For the measurements of Figs.\,\ref{fig:Evap}--\ref{fig:EvapLife}, we use two $B$-values, namely  2.43\,G and 2.55\,G, at which we perform the evaporative cooling scheme. Here we estimate $\as=109(10)\,a_0$ and $\as=134(12)\,a_0$, respectively.

\section*{Atom losses in $\Er$ and $\Dy$}
As pointed out in the main text, in the time evolution of the quantum gases in both the supersolid and the ID regime, inelastic atom losses play a crucial role. The atom losses are increased in the above mentioned regime as (i) higher densities are required so that a stabilization under quantum fluctuation effects becomes relevant and (ii) the magnetic field may need to be tune close to a FR pole to access the relevant regime of interaction parameters. (i) is at play for all magnetic species but more significant for \Er due to the smaller value of $\add$. (ii) is relevant for both $\Er$ and $\Dyo$ but conveniently avoided for $\Dy$ thanks to the special short-range properties of this isotope.

To quantify the role of these losses, we report here the evolution of the number of condensed atoms, $N$, as a function of the hold time in parallel to the phase coherent character of the density modulation observed. We count $N$ by fitting the thermal fraction of each individual image with a two-dimensional Gaussian function. To ensure that only the thermal atoms are fitted, we mask out the central region of the cloud associated with the quantum gas. Afterwards we subtract this fit from the image and perform a numerical integration of the resulting image (so called pixel count) to obtain $N$.

\begin{figure}[htbp!]
	\includegraphics[width=1\linewidth]{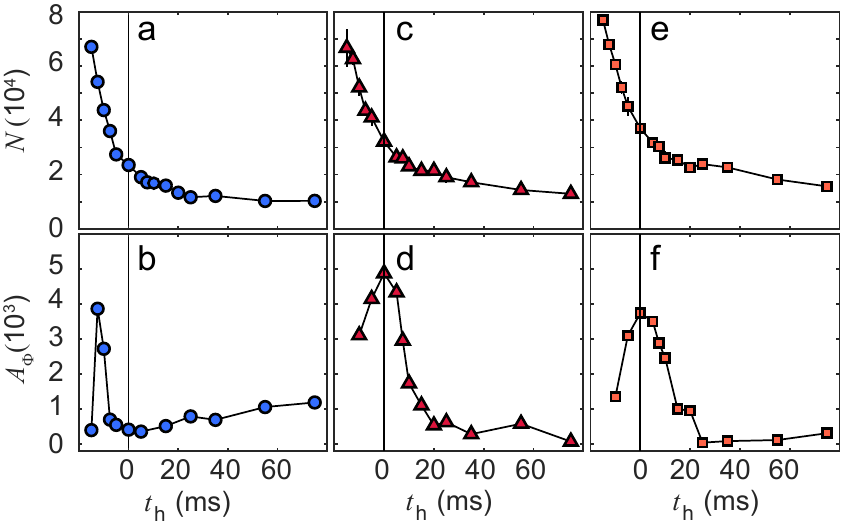}
	\caption{\label{fig:AtomLossEr}
	\textbf{atom number and coherence decays in $\Er$} Time evolution of $N$ and $A_{\Phi}$ for $\Er$ at different $a_s$, including points before $\tho=0\,$ms in the experiment. The corresponding scattering lengths are $53.3(2)\,a_0$ (a,b), $54.0(2)\,a_0$ (c,d), $54.2(2)\,a_0$ (e,f).  
	}
\end{figure}

\textit{$^{166}$Erbium -} In the Er case, a 15\,ms stabilization time is added to ensure that $\as$ is reached up to $0.3\,a_0$.
During this time, i.\,e.\,for $\tho<0$, we suspect that the time-evolution of the cloud properties is mainly dictated by the mere evolution of the scattering length. Therefore, in the main text, we report on the time evolution for $\tho\geq0$. We note that because of the narrow $\as$-range for the supersolid regime, the long stabilization time for $\as$ is crucial. However, because of the significant role of the atom losses in our system, in particular for $\Er$, the early evolution of $N$ and the cloud's properties are intimately connected. Therefore, the early time evolution at $t_h<0\,$ is certainly of high importance for our observations at $t_h\geq\,0$.

To fully report on this behavior, we show the evolution of $N$ and $A_\Phi$ during both the stabilization and the holding time in Fig.\,\ref{fig:AtomLossEr} for three different $\as$ values -- either in the ID (a,\,b) or supersolid regime (c-f). The time evolution shows significant atom loss, prominent already during the stabilization time, and levels off towards a remaining atom number at longer holding times in which we recover small BECs.
Simultaneously, in each case reported here, we observe that during the stabilization time $A_\Phi$ increases and a coherent density modulated state grows. This density modulation starts to appear at a typical atom number of $N\gtrsim 6\times 10^4$ and consecutively decays. For the lower $\as=53.3(2)\,a_0$ case, we observe that the coherent state does not survive the $\as$ stabilization time, and decays faster than the atoms loss; see Fig.\,\ref{fig:AtomLossEr}\,(a,\,b). This behavior corresponds to the ID case discussed in the main text. The central point of the present work is to identify a parameter range where the coherence of the density modulated state survives for $\tho>0\,$ and its decay time scale is similar to the one of the atom loss. In order to quantify a timescale for the atom number decay, we fit an exponential decay to $\tho\geq0\,$ms. Here we allow an offset $N_{r}$ of the fit, accounting for the BEC recovered at long holding times. In Table\,\ref{tab1Supp:AtomNumberDecay}, we report on the typical 1/10-decay times of the atom number, which are up to 50\,ms. These values are of the order as the extracted $t_\Phi$, see Table\,\ref{tab1Supp:AtomNumberDecay} and Fig.\,\ref{fig:DecayRateCoherence} of the main text. This reveals that in $\Er$ the extracted lifetime of the coherent density modulated states are mainly limited by atom loss.
\begin{center}
\begin{table}[]
\caption {Extracted 1/10-lifetime of $\Er$ atom number decay for $\tho\geq0$ and remaining atom number at long holding time for data in Fig.\,\ref{fig:AtomLossEr}.} \label{tab1Supp:AtomNumberDecay} 
\begin{tabular}{p{15mm} p{15mm} p{15mm} p{15mm}}
\hline\hline
$a_s (a_0)$  & $t_N$\,(ms) & $N_r (10^4)$ & $t_\Phi$\,(ms) \\
\hline
53.3(2)  & 32(5)   & 1.03(5)    &  -  \\
54.0(2)  & 51(9)   & 1.29(11)   &  25(6)  \\
54.2(2)  & 46(12)  & 1.7(2)     &  32(9)   \\
\hline\hline
\end{tabular}
\end{table}
\end{center}
Furthermore we note that the extracted $N_r$ values for the recovered BECs are smaller than $2\times10^4$, which is consistent with the BEC region found in the phase diagram of Fig.\,\ref{fig:theory}(f).

\begin{figure}[!htb]
	\includegraphics[width=1\linewidth]{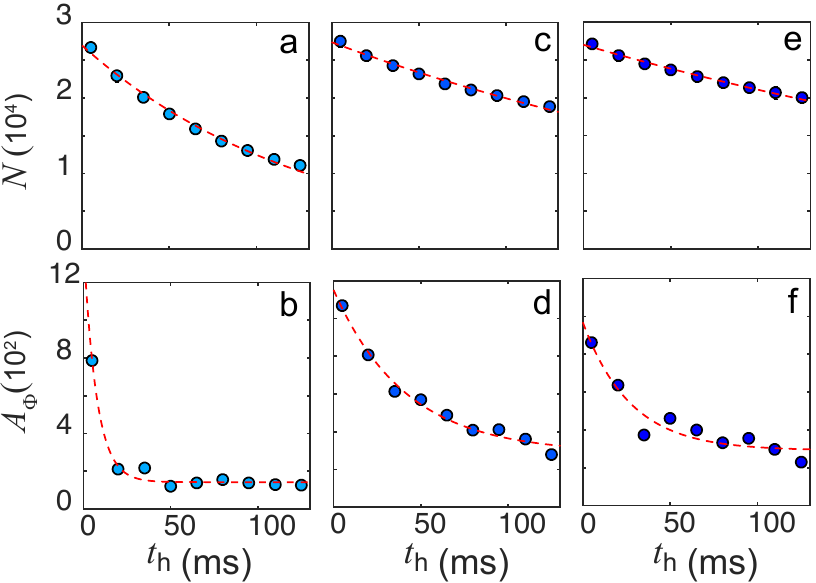}
	\caption{\label{fig:AtomLossDy}
	\textbf{atom number and coherence decays in $\Dy$} Time evolution of $N$ and $A_{\Phi}$ for $\Dy$ at different $B$ for the data of Fig.\,\ref{fig:DecayRateCoherence}. The corresponding magnetic fields are $1.8\,$G (a,b), $2.04\,$G (c,d), $2.1\,$G (e,f).  
	}
\end{figure}

\textit{$^{164}$Dysprosium -} Differently from the \Er case, for $\Dy$, we operate in a magnetic-field range in which the three-body collision coefficients are small and only moderate atom losses occur. This enables the observation of an unprecendented long-lived supersolid behavior.  To understand the effects limiting the supersolid lifetime, we study the lifetime of the condensed-atom number for different $B$. We perform this detailed study for the data of Fig.\,\ref{fig:DecayRateCoherence} of the main text, which are obtained after preparing a stable BEC and then ramping $B$ to the target value.
Fig.\,\ref{fig:AtomLossDy} shows the parallel evolution of $N$ and $A_\Phi$ for three different magnetic field values $1.8\,$G, $2.04\,$G and $2.1\,$G. Here we observe that, for all $B$ values,  $A_\Phi$ seems to decay faster than the atom number. This suggests that the lifetime of the density-modulated state in our $\Dy$ experiment is not limited by atom losses. 
To confirm this observation, we extract the 1/10 lifetimes of both $N$ and $A_{\Phi}$;  see Table\,\ref{tab:AtomNumberDecayDy}. The values confirm our observation and shows an atom number lifetime larger than $t_\Phi$ at least by a factor of $\approx 5$. In addition, we find that the ratio $t_N/t_\Phi$ varies, indicating that atom losses are not the only mechanism limiting the lifetime of the supersolid properties in Dy.

\begin{table}[htb]
\centering 
\caption {Extracted 1/10-lifetime of $\Dy$ atom number decay and $A_\Phi$ decay for data in Fig.\,\ref{fig:AtomLossDy}.} \label{tab:AtomNumberDecayDy} 
\begin{tabular}{p{15mm} p{15mm} p{15mm}}
\hline\hline
$B$\,(G)  & $t_N$\,(ms) & $t_\Phi$\,(ms)\\ \hline
1.8  & 300(12)    &  12(5)  \\
2.04  & 728(34)     &  152(13)  \\
2.1  & 926(36)      &  133(25)   \\ \hline \hline
\end{tabular}
\end{table}

\end{document}